\documentclass[a4paper,10pt]{article}
\usepackage{lineno}

\usepackage{amsmath, amsfonts, amssymb, mathrsfs}

\newcommand{\ba}{\begin{eqnarray}}
\newcommand{\ea}{\end{eqnarray}}
\newcommand{\be}{\begin{equation}}
\newcommand{\ee}{\end{equation}}

\usepackage{comment}
\usepackage{graphicx}  
\usepackage{authblk}

\title{Nanoscale Quantum Optics}
\author[1,2]{I.~D'Amico\thanks{irene.damico@york.ac.uk}}
\author[3,4]{D.~G.~Angelakis}
\author[5,6]{F.~Bussi\`eres}
\author[7]{H.~Caglayan}
\author[8]{C.~Couteau}
\author[9]{T.~Durt}
\author[10,11,12]{B.~Kolaric}
\author[13]{P.~Maletinsky}
\author[14]{W.~Pfeiffer}
\author[15]{P.~Rabl}
\author[16]{A.~Xuereb}
\author[17,18]{M.~Agio\thanks{mario.agio@uni-siegen.de}}

\affil[1]{Department of Physics, University of York  - York, United Kingdom}
\affil[2]{International Institute of Physics, Federal University of Rio Grande do Norte  - Natal, Brazil}
\affil[3]{School of Electrical and Computer Engineering, Technical University of Crete  - Chania, Greece}
\affil[4]{Centre for Quantum Technologies, National University of Singapore  - Singapore, Singapore}
\affil[5]{GAP - Quantum Technologies, University of Geneva  - Geneva, Switzerland}
\affil[6]{ID Quantique SA - Carouge, Switzerland}
\affil[7]{Faculty of Engineering and Natural Sciences, Photonics, Tampere University - Tampere, Finland}
\affil[8]{Light, Nanomaterials and Nanotechnologies-L2n, University of Technology of Troyes (UTT)  - Troyes, France}
\affil[9]{Institute Fresnel, Aix Marseille Univ, CNRS, Centrale Marseille, UMR 7249 - Marseille, France}
\affil[10]{Micro- and Nanophotonic Materials Group, University of Mons - Mons, Belgium}
\affil[11]{Institute of Physics, Center for Photonics, University of Belgrade  - Belgrade, Serbia}
\affil[12]{Old World Labs - Virginia Beach, USA}
\affil[13]{Department of Physics, University of Basel  - Basel, Switzerland}
\affil[14]{Faculty of Physics, University of Bielefeld  - Bielefeld, Germany}
\affil[15]{Atominstitut, TU Wien  - Vienna, Austria}
\affil[16]{Department of Physics, University of Malta  - Msida, Malta}
\affil[17]{Laboratory of Nano-Optics and C$\mu$, University of Siegen  - Siegen, Germany}
\affil[18]{National Institute of Optics (CNR-INO), National Research Council  - Florence, Italy}

\date{\vspace{-5ex}}

\begin{document}

\maketitle

\begin{abstract}
Nanoscale quantum optics explores quantum phenomena in nanophotonics systems for advancing fundamental knowledge in nano and quantum optics and for harnessing the laws of quantum physics in the development of new photonics-based technologies. Here, we review recent progress in the field with emphasis on four main research areas: Generation, detection, manipulation and storage of quantum states of light at the nanoscale, Nonlinearities and ultrafast processes in nanostructured media, Nanoscale quantum coherence, Cooperative effects, correlations and many-body physics tailored by strongly confined optical fields. The focus is both on basic developments and technological implications, especially for what concerns information and communication technology, sensing and metrology, and energy efficiency.
\end{abstract}

\section{Introduction}

Novel and more sophisticated technologies that exploit the laws of quantum physics form a cornerstone for our future wellbeing, economic growth and security~\cite{acin18}. Among these technologies, photonic devices have gained a prominent position because the absorption, emission, propagation and storage of light are processes that can be harnessed at the quantum level. However, the interaction of photons with single-quantum systems under ambient conditions is typically very weak and difficult to control in the solid state. Furthermore, there are quantum phenomena occurring in matter at nanometre-length scales and femtosecond timescales that are currently not well understood. These deficiencies have a direct and severe impact on creating a bridge between quantum physics and photonic devices. Nano-optics and nanophotonics precisely address the issue of controlling the interaction between a few photons and tiny amounts of matter in a large bandwidth, and the ability to efficiently funnel light down to nanoscale volumes.
Several research efforts funded by numerous national and EU projects in the last decades have already resulted into enormous progress in quantum physics and quantum optics, and in nano-optics and nanophotonics as well. Among them, the COST Action MP1403 ``Nanoscale Quantum Optics'' has been the instrument to proactively increase the interaction among the nanophotonics, quantum optics and materials science communities and to support them towards common objectives.
The grand vision of Nanoscale Quantum Optics (NQO) is the development of new ideas, novel materials, and innovative techniques to control the interaction between light and matter at will, even down to the level of individual quanta. The potential breakthroughs will have profound implications in fields as diverse as classical and quantum information processing and communication, sensing and metrology, light sources, and energy harvesting. This review follows the roadmap on NQO produced by the COST Action~\cite{roadmap17}.

\subsection{Current state of knowledge}

Research in NQO focuses on the development of novel materials (e.g. graphene, silicene, metamaterials) and optical devices (plasmonic structures, nanofibers, photonic crystals, optomechanical (OM) systems, superconducting detectors, photonic integrated circuits) that can facilitate and control strong quantum light-matter interactions, and the advancement of novel experimental methods that enable quantum degrees of freedom (color centers in solids, quantum dots (QDs), ions, molecules, neutral atoms) to be interfaced with these systems. Prominent efforts include, but are not limited to: QD devices to generate single and entangled pairs of photons on demand~\cite{senellart17}; Alternative materials for single-photon sources include organic molecules, 2D materials (e.g. hexagonal boron nitride) and thin transition-metal dichalcogenides~\cite{tran15,tonndorf17,pazzagli18}; Novel devices such as nanoscale resonant structures, plasmonic waveguides or nanofibers to achieve efficient coupling of single photons~\cite{akimov07,agio12,skoff18,baranov18}; Novel fabrication techniques for diamond- based photonics, which can be used to enhance optical coupling to individual defect color centers (such as the nitrogen vacancy (NV) and silicon vacancy (SiV))~\cite{aharonovich14,schroder16,sipahigil16}; Methods to trap cold neutral atoms to nanophotonic systems, such as tapered optical fibers, thus enabling a coherent atom-nanophotonics quantum interface~\cite{yu14,meng18}; Active exploration of metamaterials and graphene plasmonics as new platforms to channel light and control their interactions with quantum systems~\cite{koppens11,neshev18}; Chip-based OM systems have recently been cooled to their quantum ground states: coherent interactions between light and phonons have been demonstrated, creating new opportunities to use mechanical systems to manipulate quantum light~\cite{aspelmeyer14,meenehan14,leijssen17}; Superconducting devices for single-photon detection over a large wavelength range, with high time resolution, efficiency and photon number resolution~\cite{hadfield16}; Photonic integrated circuits to control the routing and interference of single photons in communications and metrology systems~\cite{sipahigil16,qiang18}; Strong light-matter coupling in semiconductor microcavities allows creation and manipulation of polariton condensates on a chip with controlled interactions~\cite{deng02,plumhof13,scafirimuto18}. Theoretical developments and experimental implementations of novel protocols to use NQO systems to generate nonclassical states and utilize these states for diverse applications in information processing, metrology, sensing, etc., are extensively studied. NQO systems are expected to have a disruptive impact, not only because of performance advantage, but also because they can be fabricated and integrated in more flexible ways to access novel parameter spaces that are not possible with macroscopic systems. For instance: NV centers have been demonstrated to serve as sensors of electric and magnetic fields with nanoscale resolution and under ambient conditions, which will have significant applications in areas such as bio and environmental sensing~\cite{degen17}; Protocols to attain single-photon nonlinear optics and sources, quantum state transfer, and hybrid coupling of disparate quantum systems, also via quantum optomechanics, are being actively investigated~\cite{maser16,desantis17,he17}; Many promising techniques to implement quantum gates for computing, entangle quantum bits, and generate nonclassical optical fields are being pursued in NQO systems such as nanoscale cavities, graphene, plasmonic waveguides, photonic integrated circuits and nanofibers~\cite{chang07b,evans18}.
There is also attention on the development of theoretical techniques to predict and quantitatively understand the rich quantum dynamics that emerges from strongly correlated quantum systems, and to understand and control the interaction of these quantum systems with complex electromagnetic environments. These efforts are critical both to position NQO systems as potentially groundbreaking quantum technologies (QT) and to leverage these systems as ``simulators'' of exotic quantum phenomena~\cite{angelakis17}.
There has been increasing interest in the exploration of quantum phenomena in biological systems (e.g. energy transport in light-harvesting complexes)~\cite{caycedosoler17}, the development of advanced experimental techniques to investigate them with high spatial and temporal resolution~\cite{hildner13}, and the investigation of bio-inspired artificial systems that exhibit quantum effects (e.g. quantum simulators).
Furthermore, advances in these research areas may have immediate technological implications and be of interest to industry. For example: Highly efficient single-photon sources on demand~\cite{quandela17}, photon-number-resolved detectors~\cite{sahin13}, and photonic integrated systems including passive power splitters and wavelength filters would have impact in the field of secure quantum communication~\cite{vlcphotonics18}; The development of switches that operate at few photon levels could drastically mitigate the high energy required for optical communication and computation in supercomputers, and it would also play a major role in the reduction of worldwide energy requirements in information \& communication technology (ICT)~\cite{kilper13}; Bio-inspired materials such as light-harvesting complexes could open new routes towards efficient photovoltaic cells~\cite{senthil18}; New approaches to single-molecule detection and quantum-enhanced measurements will expand sensor capabilities, opening new scenarios particularly within the fields of metrology, security and safety.

\subsection{Scientific and technological focus}

Three major application areas already exhibit clear evidence that the combination of quantum optics with nanophotonics is technologically valuable:
ICT, e.g. to improve single-photon sources and photon-number-resolved detectors for secure communication as well as integrated nanoscale quantum-optical solutions for ICT, Sensing \& metrology, e.g. nanosensors and quantum-enhanced measurement devices, Energy efficiency, e.g. development of new solutions for photovoltaics and energy saving. At present, several universities as well as public and private research laboratories worldwide are conducting research to introduce QT in these applications. However, the roadmap towards compact and efficient quantum devices still requires a substantial basic-research approach.
We have thus identified four research areas that deal with problems and limitations in the operation of existing QT, and that may contribute to the discovery and understanding of novel quantum phenomena for future applications: (i) Generation, detection, manipulation and storage of quantum states of light at the nanoscale; (ii) Nonlinearities and ultrafast processes in nanostructured media; (iii) Nanoscale quantum coherence; (iv) Cooperative effects, correlations and many-body physics tailored by strongly confined optical fields.
The first two priorities will also target technological aspects, such as performances and integrability of quantum photonics devices, whereas the other two include rather exploratory activities. It is worth noting that the topics involved are strongly related to three of the Key Enabling Technologies (KETs) recognized at the European level, i.e. nanotechnology, photonics and advanced materials, thus ensuring a synergic cross-KET approach to important application fields. Moreover, there is a strong overlap between the four research areas, which occurs through activities concerned with: Investigation of new materials and novel photonic structures (e.g. graphene, silicene, hybrid organic/inorganic, diamond nanostructures, plasmonics structures, semiconductor microcavities), and innovative techniques to combine them with quantum systems with a high and reproducible precision (e.g. scanning-probe techniques, two-step lithography, ion-beam milling and deposition, in-situ lithography techniques), High-throughput and low-cost fabrication methods for hybrid nanodevices and quantum photonic circuits (e.g. self- assembly, nanoimprinting, lithography), Optical methods for investigating quantum light-matter interfaces (e.g. single-molecule spectroscopy, stimulated-Raman adiabatic passage (STIRAP) and coherent population trapping), the development of novel approaches at the interface between quantum optics, nano-optics \& nanophotonics and advanced spectroscopy (e.g. nanoantenna-based spectroscopy, coherent multidimensional nanoscopy) and also modern X-ray and neutron scattering techniques; this aims at leveraging state- of-the-art experimental capabilities for new QT, Advancing theoretical techniques to quantitatively understand these phenomena (e.g. noncanonical quantization schemes, non-Markovian bath interaction models), including novel computational methods (e.g. hybrid electromagnetics/quantum mechanical algorithms or further advances in density functional theory methods). In the following, we present selected topics and discuss the various aspects concerned with each of them, offering our view on the progress that is expected/aimed at within the next 5 years.


\section{Generation, detection, manipulation and storage of quantum states of light at the nanoscale}

\begin{figure}[h!]
\centering
\includegraphics[width=0.6\columnwidth]{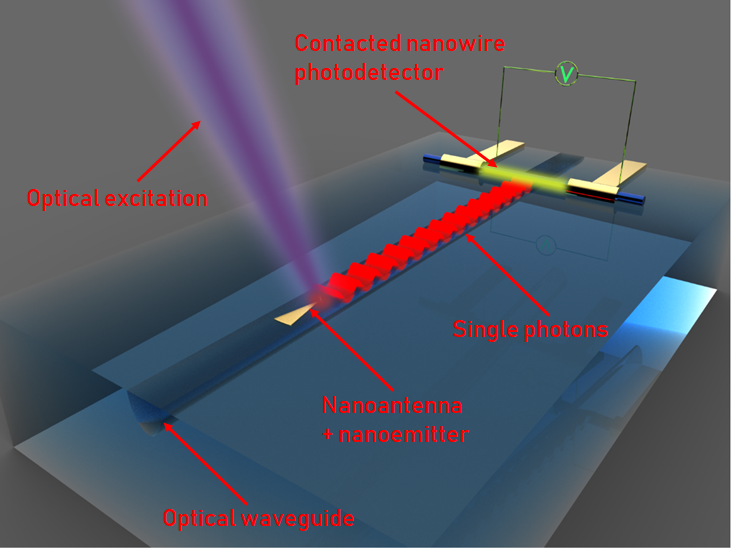}
\caption{Schematic view of a fully integrated quantum circuit with an optical waveguide guiding single photons emitted efficiently from a nanoemitter coupled to a nanoantenna. The photons are then sent to a nanosystem photosensor for detection.\label{fig:Circuit}}
\end{figure}

NQO aims at developing integrated photonic structures for the generation, detection, manipulation and storage of quantum states of light that involve quantum emitters,  waveguides, cavities, junctions and detectors. For instance, NQO must enable devices with more than 90\% (more than 99\%, ideally) emission from a quantum emitter into an optical fiber. Another important goal is to be able to probe and to collect light from a single nanostructure via an optical communication bus for reversible interaction (i.e., photon-to-quantum-emitter and quantum-emitter-to-photon). Figure~\ref{fig:Circuit} illustrates how a quantum integrated circuit at the nanoscale would look like, with an optical waveguide (made of ion-exchange in glass for instance~\cite{madrigal16}), a quantum emitter coupled to a nanoantenna, excited optically (or electrically in the future) and single photons waveguided towards a nanowire photodetector (made of superconducting material for instance). 

\subsection{Single-photon sources}

Single-photon sources are at the heart of many applications in quantum technologies as an essential building block. To date, producing a scalable source of on-demand and efficient indistinguishable photons remains a challenge~\cite{lounis05,takeuchi14,aharonovich16}. Several systems are being under investigation ranging from single trapped atoms to artificial atoms (QDs) or isolated defects in high band-gap semiconductors such as diamond. Microwave photons are also currently produced in superconducting platforms although visible to near infrared photons are still more desirable for interfacing with telecom bandwidths or for silicon detectors. Within this context, nanosources of light seem the best potential option to meet these criteria, but many key issues are still present. As knowledge and means of controlling single-photon sources develop, more fundamental questions arise, in particular in terms of coherence, light-matter coupling and many- photon interaction. As such, novel experimental findings require progress in theoretical tools such as: Finite difference time-domain simulation, Optical Bloch and Maxwell-Bloch equations, density matrix approach for coupling with reservoirs, modelling of light extraction and collection or density functional theory.
By the same token, in the last few decades, a plethora of nanoscale single emitters have emerged, each with their pros and cons. The quest for the best host material and the best quantum emitter is still under way. Many materials are currently explored (semiconductors, diamond, 2D materials, etc.) as well as different geometries (rod, dots, pillars, wires, etc.), and even different dimensions (3D confinement, defects in 2D materials, etc.). Moreover, efforts in material engineering are important to attain scalable and reproducible single-photon sources. Examples of these systems include: III-V semiconductor self-assembled QDs emitting at 900 nm, 1.3 $\mu$m and 1.55 $\mu$m, organic molecules, color centers in diamond, 1D materials, such as nanowires or carbon nanotubes, 2D materials, such as hexagonal boron nitride or thin transition-metal dichalcogenides but also spontaneous parametric down conversion and spontaneous four-wave mixing in micro- and nanostructures. We can also cite rare earth ion doped crystals and finally mention the important work towards full integration of electrically pumped systems (e.g. QDs, color centers in diamond).
In terms of challenges, electrically controlled single photon sources, which work under ambient conditions and emit identical photons, remain an important goal. In the not-so-long run, single-photon sources serving different kinds of applications will be on the market (some already are, albeit not turnkey) and, as such, they will have to be ready for commercial challenges in terms of performances, cost and reliability. Many points will then have to be addressed, such as: standardisation and calibration of single-photon sources~\cite{rodiek17}, scalable fabrication of indistinguishable single-photon sources and sources of entangled photon pairs~\cite{schwartz16,somaschi16}, but also development of stand-alone and fiber-coupled single-photon sources and efficient electrical pumping schemes~\cite{fedyanin16}.

\subsection{Superconducting single-photon detectors}

\begin{figure}[h!]
\begin{center}
\includegraphics[width=0.4\columnwidth]{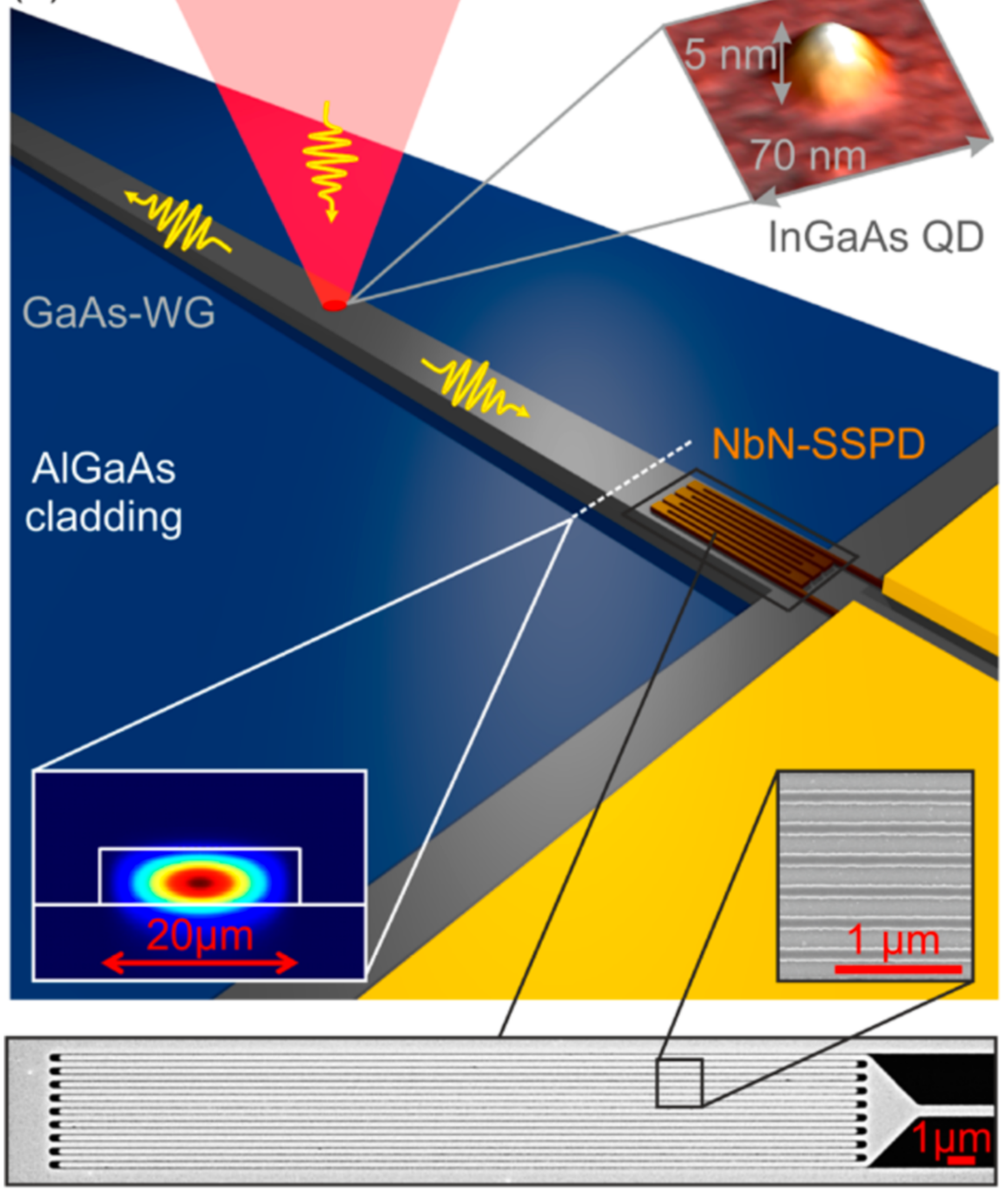}
\caption{Integration of a QD single-photon emitter with on-chip superconducting single-photon detector. Schematic of a 2-mm-long GaAs ridge waveguide containing a single layer of InGaAs QDs. The dots are excited by a free-space pulsed laser diode and emit in the waveguide. The waveguide then propagates light towards an evanescently coupled NbN superconducting single-photon detector (reprinted Fig.~1 with permission from Ref.~\cite{reithmaier15}. Copyright (2015) by the American Chemical Society).}
\label{fig:photoncircuit}
\end{center}
\end{figure}

In one big potential scheme of what integrated quantum optics/quantum photonics and quantum nanodevices would look like (see Fig.~\ref{fig:photoncircuit}), one needs to create photons, to propagate them and to detect them.
 
Like single-photon sources, single-photon detectors are indispensable for applications in QT~\cite{hadfield09}. In the case of detectors, the challenge of detecting a single photon with close-to-unity efficiency over a wide wavelength range remains~\cite{marsili13}. Moreover, these detectors should be able to resolve photon number and ideally work under relatively relaxed conditions (e.g.~turnkey operations). Ideally, one would also work at room temperature, but working at cryogenic temperature is not so much of an issue with the recent progresses in cryogenics and in particular using closed-cycle cryostats, which do not require a constant supply of helium. Superconducting nanowire detectors seem the best potential option to meet this challenge, but there are still many associated problems to be solved~\cite{engel15,natarajan12}. Again, as knowledge and means of fabricating and engineering these single-photon detectors develop, fundamental questions arise, especially in terms of efficiency, speed or insertion into more complex photonic circuitry. Moreover, studying the detection mechanism allows us to determine the physical limits to the operation of these devices. It is at present not fully clear how far away we are from these limits. Theoretical work is thus necessary for a better and more fundamental understanding of the systems, such as understanding superconductivity in nanostructures, as well as numerical simulations of the detection mechanism, study of the superconducting single-photon detector (SSPD) mechanism in unconventional superconductors to establish the feasibility of high-critical-temperature (Tc) SSPDs~\cite{hadfield16}, but also understanding non-Bardeen-Cooper- Schrieffer (BCS) electrodynamics (e.g. interplay of
disorder, interaction and photon-driven out-of-equilibrium superconductivity). For such technologies, material engineering is of prime importance for superior single-photon detectors. Many new materials have emerged recently as potential candidates. Meanwhile, progress in nanofabrication enables different paradigms for controlling detection at the nanoscale. Examples of topics being explored in this area include superconducting nanowire single-photon detectors (meanders, nanodetectors), photon-number-resolving transition edge sensors (TES), well-established and new superconducting materials: NbN, NbTiN, WSi, NbSi, MoSi, YBCO, InO. Design is also paramount with multilayers and new geometries occur such as waveguide detectors~\cite{najafi15,ferrari18}, kinetic inductance detectors or using alternative substrates. We should also note that optical, electrical, temperature and magnetic field studies at $< 1$~K have to be done but also high-temperature operation. We should also stress that semiconductor photodetectors are still around and still widely used. Performances of such detectors whether being in silicon in the visible to near infra-red or being in germanium or AlGaAs for infra-red detection are getting better and material issues are tackled with better specifications. Such materials are worth developing for applications and active research is still carried out on these topics.
An ideal single-photon detector has yet to be developed, where all the photons are sorted and detected, possibly over a large energy range (down to microwave region for some applications). Like single photon sources, single-photon detectors are already on the market. However, there are still challenges in terms of performances, cost and reliability that need to be addressed and here is a non-exhaustive list of issues such as standardisation and calibration of single-photon detectors, some sort of specification sheets and development of affordable, small-footprint cryocoolers for commercial systems.

\subsection{Manipulation and storage of single photons}

\begin{figure}[h!]
\centering
\includegraphics[width=0.8\columnwidth]{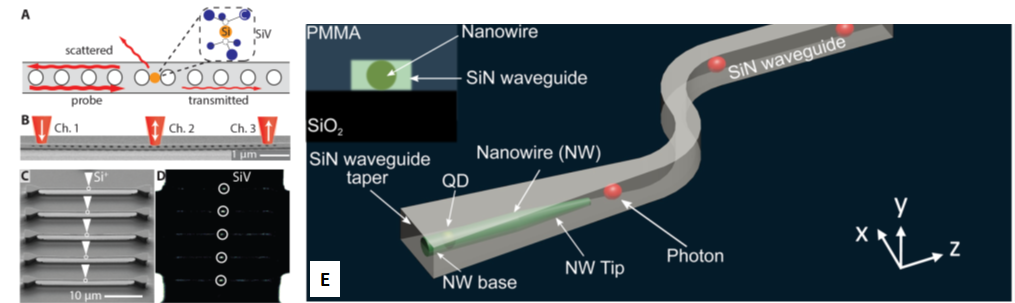}
\caption{(a) Schematic of a single SiV color center in a diamond photonic cavity. (b) Scanning electron micrograph (SEM) of such a diamond photonic crystal. (c) SEM of diamond cavities fabricated without color centers. (d) SiV fluorescence for each nanocavity after focused Si$^+$ ion beam implantation to create the color centers. (reprinted Fig.~1a-d with permission from Ref.~\cite{sipahigil16}. Copyright (2016) by the American Association for the Advancement of Science). (e) Schematic view of a III-V QD in a nanowire embedded in a Si$_3$N$_4$ photonic waveguide after micro-manipulation (reprinted Fig.~1a with permission from Ref.~\cite{zadeh16}. Copyright (2016) by the American Chemical Society).\label{fig:manipulation}}
\end{figure}

Besides generation and detection of single photons, the manipulation and/or storage of quantum states of light must also be implemented in integrated circuits~\cite{sangouard11}. Moreover, scaling quantum photonic technologies require efficient and long-term single-photon storage~\cite{kennard13}. A number of requirements are important, such as storage and retrieval efficiency, storage time and reduction of background noise to reach a high fidelity. These and other aspects have been actively studied for many years such that complex circuitry is now available and manipulation and/or storage are improving towards practical applications.
Manipulation of single photons has now reached a high degree of control and scalability compared to storage. The latter is more challenging, as it requires finer control over light-matter interaction. Therefore, it is in this area that a great deal of theoretical work is needed. Examples of key issues include: semiclassical description of (arrays of) multilevel atoms with optical near fields, full quantum description of the interaction of multilevel atoms with resonator modes but also schemes for the manipulation and storage of quantum states of light on a photonic chip and designs for coupling single photons to low-loss photonic integrated circuits.
Even though manipulation of multiple single photons has been achieved~\cite{bell14} and is scalable to some extent, there are still challenges for several schemes in quantum information processing. Likewise, quantum communication demands a stringent level of control that has yet to be achieved~\cite{barz15}. Quantum storage of single photons preserving coherence and other intrinsic properties is definitely very challenging~\cite{bussieres13}, and different materials and systems are currently investigated, such as cavity quantum electrodynamics (CQED) with single emitters coupled to micro- and nanoresonators, quantum memories based on ion-doped crystals, nonlinear waveguides (e.g. periodically poled lithium niobate), integration of solid-state single-photon source (e.g. QD nanocrystals) in optical microresonators and
in quantum photonic circuits, tapered optical fibers with a nanosize waist and >99 \% transmission~\cite{sayrin15}.
The performance of single photon manipulation has greatly increased recently, especially with the use of integrated optics components. Like in microelectronics, we can classify physical systems into two types, the top-down and the bottom-up approaches. The top-down approach consists essentially into two platforms, which are III-V semiconductor QDs and the diamond platform. Top-down means that one starts from a bulk or from semi-bulk material, which is etched down to the desired structure. For epitaxial QDs in semiconductors (usually III-V materials), the idea is to grow QDs first and then etch down the material in order to build a photonic device around it. Nowadays, structures are made, where a single QD, electrically driven, is embedded in a micropillar with Bragg mirrors in order to benefit from CQED effects. A second promising platform has emerged, which consists in making nanophotonic structures in diamond in order to more effectively interrogate single defects (NV or SiV centers). Figure~\ref{fig:manipulation}a-d present an example of the degree of control for fabricating and positioning color centers in diamond photonic structures~\cite{sipahigil16}. Along the same line, one can mix nanodiamonds into a photoresist, which can then be ‘carved’ by two-photon photolithography in order to obtain photonic structures such as ring resonators or more complex structures.
The second approach is the so-called bottom-up, which consists of combining together photonics nanocomponents in order to make a more complex structure with a source of light, an optical circuitry and eventually a detector. For example, a single NV center in a nanodiamond is inserted into a V-groove in gold, coupling light from quantum emitter to the plasmonic modes of the V-groove~\cite{bermudez15}. Colloidal CdSe/ZnS QDs have been used sandwiched between two high-refractive-index materials, but also using DBT molecules in anthracene next to a Si$_3$N$_4$ waveguide coupled together by a plasmonic resonance due to the presence of a thin film of gold~\cite{kewes16}. One can also make use of epitaxial growth of a single QD within a nanowire being already a nice photonic device that can be micromanipulated and embedded into another waveguide (see Fig.~\ref{fig:manipulation}e from Ref.~\cite{zadeh16} as an example). If we want to be exhaustive, we should mention some fascinating and challenging results going forward for coupling atomic systems with nanophotonics where a single atom is trapped in the near field of a nanoscale photonic crystal cavity realising a strong photon-atom coupling, where one modifies the other one’s phase~\cite{tiecke14}. In the same family of experiments, let us mention the idea to trap atoms within the evanescent field of a (nano)fiber and/or a resonator~\cite{volz14}.

There are more complex hybrid systems now appearing, such as the coupling of entangled photons emitted from III-V QDs with an atomic vapour~\cite{trotta16} and metamaterial systems that can help harnessing light-matter interaction at the nanoscale~\cite{altuzarra17}. Nevertheless, there is still need for improvement to turn devices into products. For the storage of single photons, it is even more challenging to achieve commercial performances (specific materials, low temperatures, etc.), e.g. for quantum repeaters, and efforts need to be focused in these areas for developing strong light-matter interactions at room temperature, low-loss and low-cost coupling to optical fibers and low-loss optical channels down to 1 dB/km.




\section{Nonlinearities and ultrafast processes in nanostructured media}

Optical nonlinearities and ultrafast dynamics are ultimately intertwined in nanoscale environments. This is readily seen by considering for example electron velocities in materials and the nanometer length scale. Electron propagation over 10 nm at a typical Fermi velocity of 10$^6$ m/s happens within 10 fs. Studying these dynamics directly in the time and spatial domain involves the methods of present state ultrafast laser technology providing ultrashort optical pulses in a wide range of the electromagnetic spectrum. Enabled by the intense optical fields and the local field enhancements encountered in nanoscale structures optical nonlinearities play an important role and provide means to attain new quantum optical functionalities. Studying the rich phenomenology of nonlinear optics on the nanoscale formed the first playground for ultrafast nanoscale quantum optics and is presented more thoroughly in Section~\ref{sec:nanonlresp}. The design, fabrication, and investigation of broadband interfaces, i.e. antennas and emitters, between plane wave far-fields and optical near-fields that are needed to achieve ultrafast operation of nanoscale photonic devices forms an important field of research (Section~\ref{sec:nanoemant}). Energy and charge transport in nanostructures deviates significantly from mechanisms applicable in large scale structures. 
The related spatio-temporal dynamics are of interest, being essential in many real-world applications such as for example light harvesting (Section~\ref{sec:nanotrans}). The ultrashort time and length scale at which the relevant processes occur pose the challenge to develop new types of spectroscopy that provide simultaneous ultimate spatial and temporal resolution (Section~\ref{sec:specnano}). Quantum functionality in nanoscale systems offers routes to miniaturize and highly integrate quantum technologies. One key element is employing nonlinear processes for active nanophotonics enabling for example nanoscale lasers and amplification schemes to overcome intrinsic damping (Section~\ref{sec:actnano}). More complex quantum functionalities are conceivable if the spatio-temporal evolution of nanoscale systems is coherently controlled (Section~\ref{sec:ccnano}).  

\subsection{Nanoscale nonlinear response}
\label{sec:nanonlresp}

\begin{figure}[h!]
\center\includegraphics[width=0.8\columnwidth]{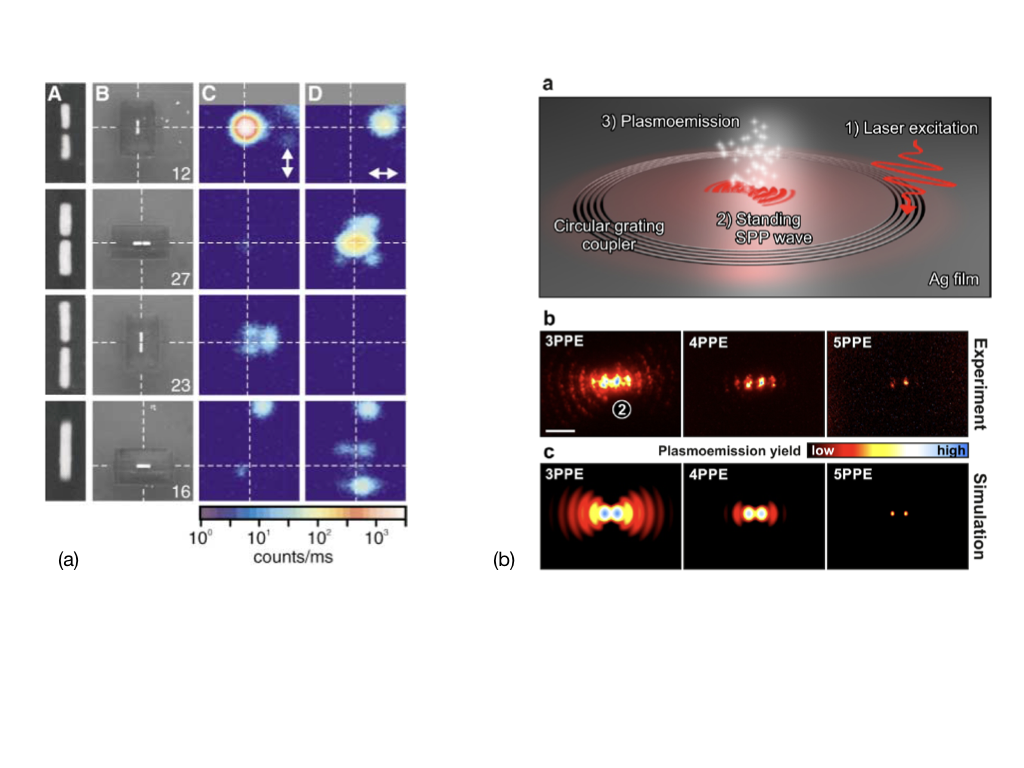}
\vspace{-2cm}
\caption{Nonlinear processes in nanoplasmonics. (a) Demonstration of the resonant interaction between optical nanoantennas shown as SEM micrographs in A and B. First and second row show confocal images of resonant white light generation for two incident polarizations indicated by white arrows, whereas third and fourth row show the non-resonant case. In the resonant case the nonlinear yield is enhanced by about 4 orders of magnitude (reprinted Fig.~1 with permission from Ref.~\cite{muehlschlegel05}. Copyright (2005) by the American Association for the Advancement of Science). (b) The upper part visualizes the generation of focusing surface plasmon polaritons in a circular plasmon corral that leads to a multi-plasmon electron emission from the center of the corral. The lower two rows show measured and simulated emission patterns for different order multi-plasmon electron emission channels. (reprinted Fig. 3 with permission from Ref.~\cite{podbiel17}. Copyright (2017) by the American Chemical Society).\label{fig:nanonlresp}}
\end{figure}

The nonlinear response is typically only a weak effect in bulk materials or interfaces. However, the field concentration achieved in nanophotonics enhances the signal from nanoscale objects. For example, as shown in Fig.~\ref{fig:nanonlresp}a the early demonstration of optical nanoantennas employed nonlinear optical processes, i.e., white light generation in antenna nanogaps, to demonstrate the resonant character of light antenna interaction~\cite{muehlschlegel05}. The nonlinear effect is here employed to emphasize the resonant character of the interaction since the third order nonlinearity enhances the signal contrast between resonant and non-resonant interaction by orders of magnitude. However, nonlinear processes are essential for functionality such as switching operations~\cite{macdonald09} or frequency conversion.
Two aspects are of particular interest in NQO: (i) enhanced nonlinear response in optical near-fields and (ii) nonlinear CQED at the few- or single- quantum level. In contrast to bulk nonlinear materials, the  field distribution can no longer be envisioned as homogeneous, which has generated renewed interest in the actual mechanisms of the nonlocal~\cite{raza15} and nonlinear~\cite{niesler09} response on a microscopic scale. The investigated systems (metal or hybrid nanostructures) are still too large for a full \textit{ab-initio} quantum theoretical treatment. Thus modelling relies on the combination of Maxwell solvers for complex geometries with implemented classical~\cite{moeferdt18} or quantum models~\cite{marinica12} yielding the local linear and nonlinear responses. In terms of experimental achievements, the capability to measure the response from individual nanostructures is essential. This avoids inhomogeneous line- broadening effects and dispersion in the spatial distribution of measured signals. The huge nonlinearities observed in conventional CQED occur at the few- or single-quantum level~\cite{gleyzes07} and form a blueprint for the development of nanoscale quantum-optical devices~\cite{torma15}. Strong coupling with plasmons has been demonstrated for ensembles of quantum emitters~\cite{vasa13} and between quantum emitters randomly placed in a nanophotonic cavity, for example~\cite{kim11}. Based on even higher field enhancements, strong coupling at room temperature has recently been achieved for a single dye molecule placed in a plasmonic gap resonance~\cite{chikkaraddy16}. To increase the coupling strength the quality factor of nanophotonic cavities must be improved. This will also extend the time window for coherent manipulation of embedded quantum systems. For a broader discussion of nonlinear effects arising from nanoscale strong coupling phenomena refer to the recent review by Vasa et al.~\cite{vasa18}. Beyond realization of CQED schemes also the quantum character of the plasmon excitation often employed in NQO is of rising interest. Whereas it is mostly conceived as a classical entity there are recent examples that demonstrate that the nonlinearity of the collective plasmon excitation itself plays a significant role in multi-quantum excitation processes~\cite{podbiel17} (see Fig.~\ref{fig:nanonlresp}b).

Investigation of the nonlinear response relies on the close interaction of experiment and theory. Experimentally determined nonlinear signals preferentially from a single nanostructure are quantitatively matched to theoretical simulations~\cite{decorny16}. Highly nonlinear optical materials embedded in nanophotonic devices open a route to nanoscale all-optical switching and coherent photon-photon interactions. Adaptation of CQED for nanophotonics seems straightforward. However, the presence of strong loss channels poses severe restrictions and robust, and ultrafast, coherent manipulation protocols need to be developed. There are thus many aspects that require further consideration from a theoretical point of view, especially: (i) integration of quantum mechanical response in Maxwell solvers, (ii) nonlinear response in nanostructures, (iii) ultrafast nanoscale all-optical switching, (iv) nanoscale CQED based on Jaynes-Cummings and Tavis-Cummings models, (v) the role of open quantum systems, and (vi) strategies enabling nanoscale strong and ultrastrong coupling with single emitters. 
The concept of nonlinear response in nanostructures is currently investigated for plasmonic systems with rather simple hydrodynamic and jellium models. Application to anisotropic dielectrics and hybrid nanostructures and a more refined treatment of the quantum system response are needed. Nanoscale CQED relies on resonators with tiny mode volumes and thus requires a highly controlled embedding of quantum emitters where atomic scale conditions become essential. Improving the quality factor of nanoscale cavities would facilitate reaching the CQED regime and strong or ultrastrong coupling, but it would limit the bandwidth of the cavity on the other hand. Single-molecule spectroscopy methods are employed to study CQED effects. Key focus areas for expanding the methodology in this field of research cover various novel materials, hybrid systems and innovative experimental techniques. This includes in particular: (i) design of advanced plasmonic and hybrid nanostructures, (ii) exploitation of low-dimensional materials (e.g. nanowire, nanotubes, graphene) and their hybrid structures, (iii) development of low loss nanophotonic wave guides, nano cavities, and nonlinear metasurfaces, (v) controlled embedding of quantum emitters (defect centers in diamond, QDs, molecules, etc.), (vi) implementation of advanced coherent single-molecule spectroscopy methods based on second- and third-harmonic generation or tip-enhanced nonlinear spectroscopies.
Improved understanding of optical nonlinearities could open routes to further enhanced nonlinear responses from nanoscale photonic devices. Hybrid nanostructures and tailored fields and responses could enable ultrafast all-optical switching. Reaching the strong coupling regime in nanoscale CQED requires cavities with huge local field enhancements, tiny mode volumes, and advanced assembly methods that allow controlled embedding of quantum emitters. Further investigations and improvements are therefore needed, for example, in: (i) probing the nonlinear local response of nanostructures with high spatial resolution, (ii) studies of the nonlinear response in hybrid nanostructures, (iii) a full quantum theory of linear and nonlinear response in complex geometries, (iv) development of ultrafast all-optical switching in nanophotonic devices, (v) reaching ultrahigh local density of photon states in nanostructures, (vi) development of nanoscale quantum emitters with stable spectral properties, (vii) single- or few-quantum nonlinearities in nanoscale devices, and (viii) ultrafast CQED.

\subsection{Nanoemitters and nanoantennas}
\label{sec:nanoemant}

\begin{figure}[h!]
\center\includegraphics[width=\columnwidth]{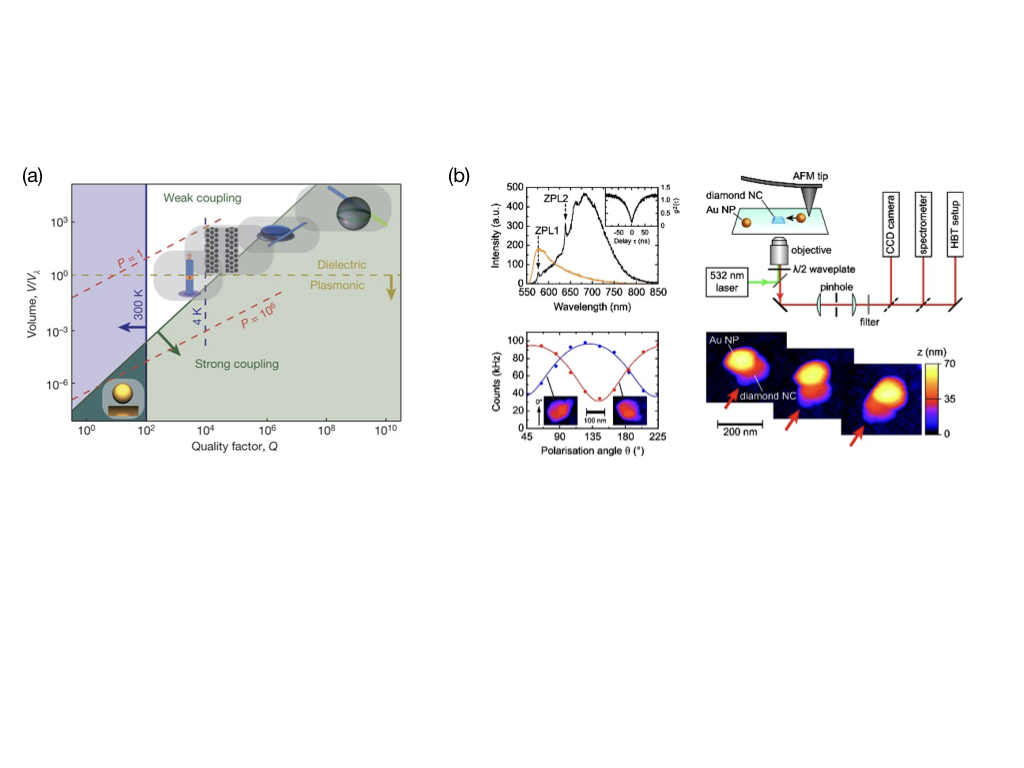}
\vspace{-5cm}
\caption{(a) The quality factor, Q, of nanoanteannas is plotted against its effective volume, $V/V_\lambda$ showing strong-coupling (green arrow), room-temperature (blue arrow), and plasmonic (orange arrow) regimes for single emitters (reprinted Fig.~1a with permission from Ref.~\cite{chikkaraddy16}. Copyright (2016) by the Springer Nature Publishing Group). (b) Optical properties and nanoscale manipulation of diamond nanocrystals and gold nanoparticles. The gold sphere is moved around the diamond nanocrystal, which stays fixed in its orientation, thus testing the optimal configuration for emission enhancement (reprinted Fig.~1 with permission from Ref.~\cite{schietinger09}. Copyright (2009) by the American Chemical Society).\label{fig:nanoemant}}
\end{figure}

Nanoemitters and nanoantennas play an essential role in nanoscale quantum optical phenomena due to their ability to link the nanoworld to the far-field transverse electromagnetic radiation. The nanoantennas operate in a similar way to radio antennas but at higher frequencies and bridge the size and impedance mismatch between nanoemitters and free space radiation, as well as manipulate light on the scale smaller than wavelength of light~\cite{agio13}. 
The wave nature of light limits the efficient guiding structures to the (sub)wavelength dimensions. Nanoantennas overcome these constraints, allowing unprecedented control of light-matter interactions within subwavelength volumes~\cite{anger06,kuehn06a}. Such properties have attracted much interest lately, due to the implications they have both in fundamental research and in technological applications.  Broadband response and directionality~\cite{taminiau08a,kuehn08} were demonstrated and the interfacing to nanoscale transmission lines~\cite{rewitz12} and cavities is presently investigated~\cite{schoen13}. Nanoantennas couple to the electromagnetic fields emitted by molecules, atoms, or QDs placed close by, leading in turn to a strong modification of the radiative and nonradiative properties of the emitter. 
As depicted in Fig.~\ref{fig:nanoemant}a both effective mode volume and the quality factor of the involved electromagnetic modes determine the achieved coupling strength. In contrast to conventional CQED this allows reaching the strong coupling limit also for rather low-quality resonators. However, strong coupling to single quantum emitters is challenging~\cite{chikkaraddy16} and even advanced methods of nanofabrication do not provide a satisfying level of control. On the other hand, self-organized assembly of nanoantenna-quantum emitter structures may provide sufficient control~\cite{pfeiffer12}. Other strategies employ scanning probe manipulation of plasmonic structures to optimize and tailor the quantum emitter-antenna interaction (see Fig.~\ref{fig:nanoemant}b, Ref.~\cite{schietinger09})

Photons play a key role as information carriers in quantum technologies, encoding and communicating quantum information in various degrees of freedom (polarization, wavelength, timing, or path). Typically, these technologies require strong interaction between weak single photon fields and matter. To boost this interaction, nanoantennas can be used to tailor the optical field to strongly interact with the nanoemitter~\cite{agio13}.
For quantum optical application not only, the strong coupling of local excitations to the far field is of interest but also efficiency of energy transfer.  It is beneficial to minimize the dissipation to the dark modes or purely evanescent modes for an efficient coupling of quantum emitters. To achieve this, combination of different modes and mode hybridization can be used for tailoring the response of the system via Fano resonances~\cite{lukyanchuk10} and mode hybridization~\cite{prodan03}. Coupling to cavities and transmission lines is essential in the realization of next-generation photonic circuits and quantum information processing~\cite{sato11}. Sub-radiative and super-radiative modes are employed to adapt the properties of nanoantennas for specific quantum functionalities. The generation and control of such modes may extend the interactions, which is typically limited by the cavity size~\cite{solano17}.
Up to now, research focuses mostly on metallic, i.e. plasmonic, nanostructures serving as nanoantennas and nanoemitters. Although the losses induced by material imperfections can be minimized by employing single crystalline materials, intrinsic losses remain large. Dielectric and hybrid materials could serve to minimize these losses further. A number of strategies for loss mitigation (or avoidance) are being pursued. These involve using advances in fabrication aimed at reduction of surface roughness and effects, using highly doped semiconductors in place of metals, using polar dielectrics in the \emph{Reststrahlen} region (phononics), and using media with optical gain. 
Dissipation and decoherence in nanophotonic systems is one of challenges that is limiting the use of nanoantennas for quantum optical applications. If the coupling to long lived dark modes can be controlled and minimized employing dark modes, these nanoantenans can be used as an efficient dissipation channel for coupling to the far field. Some of the other challenges in these nanophotonics systems are  precise manipulation and placement of individual quantum emitters in the vicinity of optical nanoantennas. The two main challenges in developing nanophotonic structures are controlling the relative position and orientation of the quantum emitter with respect to the structure and the stability of the emitters. Some of the approaches that have been successfully used to position quantum emitters close to plasmonic structures is to chemical conjugate~\cite{ringler08} or using atomic force microscopy (AFM) nanomanipulation~\cite{schietinger09} (see Fig.~\ref{fig:nanoemant}b). Moreover, self-organized assembly of nanoantenna-quantum emitter structures and precise positioning/growth of quantum emitters has great potential for the field of nanophotonics. 
Recent development on the integrated plasmonic circuits containing quantum emitters to build quantum information processing devices provides successful implementation of nanoscale quantum optics for ICT, sensing \& metrology, energy efficiency. 

\subsection{Nanoscale transport}
\label{sec:nanotrans}

\begin{figure}[h!]
\center\includegraphics[width=\columnwidth]{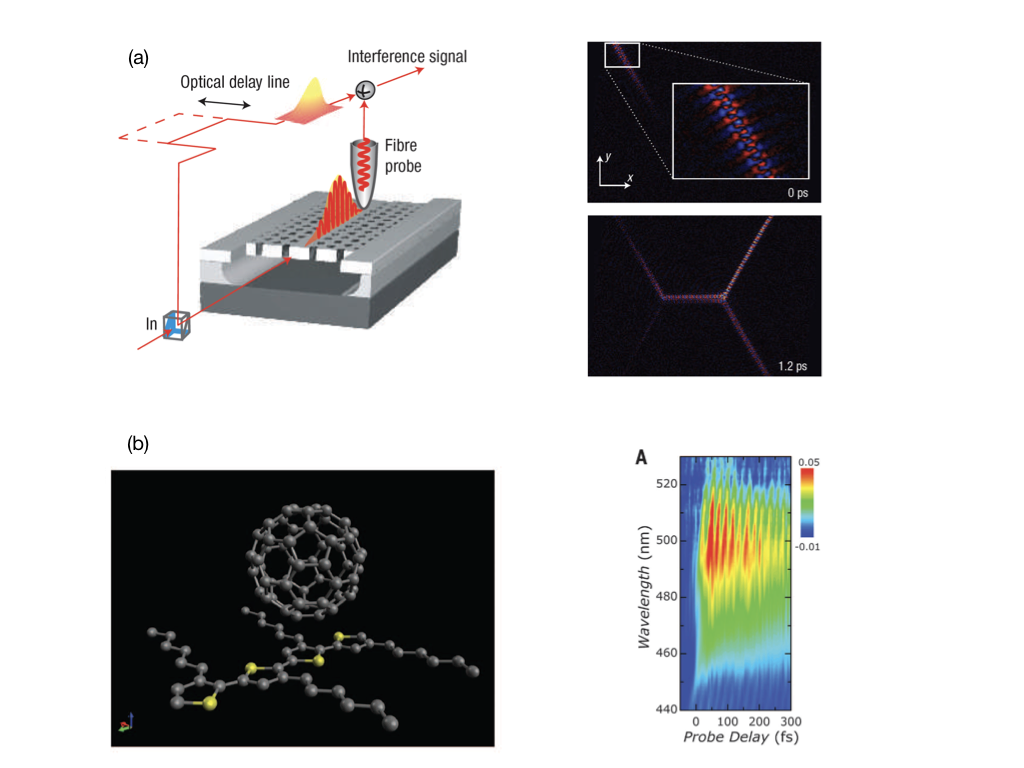}
\caption{Spatiotemporally resolved investigation of energy and charge transfer. (a) Recording of pulse propagation in a nanophotonic wave guide structure based on spectral interference measurements resolving the nanoscale electrical field evolution (reprinted Fig. 2 and Fig. 3 with permission from Ref.~\cite{engelen07}. Copyright (2007) by by Springer Nature Publishing Group). (b) Charge transfer dynamics in the P3HT:PCBM blend. The left part shows the Molecular structure of the P3HT:C60 compound. In the right panel the transient differential absorption signal reveals the photon induced charge transfer from the the polymer to the fullerene that is facilitated by the coherent oscillations seen in the transient signal. (reprinted Fig. 1c and Fig. 2a with permission from Ref.~\cite{falke14}. Copyright (2014) by the American Association for the Advancement of Science).\label{fig:nanotrans}}
\end{figure}

Nanoscale electron, spin and energy transport occurs intrinsically on an ultrafast timescale and this poses a challenge for present state methodology. In general simultaneous spatial and temporal resolution on the corresponding length and time scales is required to in detail resolve the basic mechanisms. Energy transfer relies on coupling and competes with other relaxation channels. Hence efficient energy transfer has to be fast~\cite{sundstrom00}. Although very efficient nanoscale energy transport is for example attained in natural photosynthetic light harvesting, its implementation in artificial light harvesting is still in an early stage~\cite{bredas16}. Time-resolved nonlinear spectromicroscopy serves as a valuable tool to visualize coherent energy transport occurring in surface plasmon polaritons~\cite{bauer02} and more complex electromagnetic modes. In many cases the actual spatial transport effect is only inferred indirectly from the spectroscopic signal since the method does not provide spatial resolution. By employing methods with high spatial resolution this can be overcome and for example scanning probe methods are combined with time-resolved techniques as it is shown in Fig.~\ref{fig:nanotrans}a. Here a scanning near field optical microscope (SNOM) tip is used to pick up the local field in a nanophotonic waveguide and spectral interference with a coherently related reference pulse is used to retrieve the full spatio-temporal evolution of the field~\cite{engelen07}. Such methodological combinations are not limited to all optical signals. For example photoelectron electron microscopy in combination with pulsed excitation serves to investigate nanoscale coherent transport of energy~\cite{aeschlimann17}. 
The theoretical treatment of energy transfer in optical nanostructures is based on Maxwell’s equations. Coupled localized plasmon polaritons, propagating plasmonic modes, and F\"orster energy transfer between quantum emitters are commonly applied concepts. Still in more complex systems the role of coherence and decoherence in the whole process is still controversially discussed and further investigations are required. This is intimately linked to the fact that for systems in which quantum emitters are embedded quantum optical models have to be developed and employed.  
Compared to energy transport the study of charge transport on short scale poses even larger challenges. Ballistic charge transport is limited to the mean free path of the carriers. Typically it is restricted from a few nanometers to about 10 nm for excited carriers close to the Fermi level. Thus lateral transfer is difficult to directly measure. However, transient charging after photoexcitation~\cite{pfeiffer04}, induced photocurrents in heterostructures~\cite{differt12}, exciton microscopy~\cite{yoon16}, and photoemission of excited electrons provide access to nanoscale charge transport phenomena. Ultrafast electron transport is significant in the ultrafast demagnetization observed in ferromagnets after pulsed excitation~\cite{eschenlohr13}. Complex media and composite materials as they are used for example in organic photovoltaics are based on nanoscale heterogeneity and charge transport on the corresponding length scales. Investigation of charge separation and theoretical modeling in rather large such systems has now been demonstrated~\cite{falke14} (see Fig.~\ref{fig:nanotrans}b).

Excited electron transport is limited by inelastic electron collisions treated phenomenologically using the Boltzmann equation. On short time and length scales ballistic transport dominates. Thus, the development of nano-optical tools for the investigation of transport in nanomaterials and quantum systems involves aspects as diverse as surface plasmon polaritons, F\"orster energy transfer, excited electron relaxation, nonequilibrium charge dynamics in nanostructures, Boltzmann equation in complex geometries, and ballistic transport in constrained geometries.
For further investigations at the very small length scale and simultaneous very short time scale novel time-resolved imaging techniques are essential. Already demonstrated are time-resolved SNOM and time-resolved PEEM. Completely new possibilities arise with the development of ultrafast electron microscopy. Ultrashort and highly focused electron pulses can probe electromagnetic fields~\cite{yurtsever12} and free space electron propagation~\cite{vogelsang18}. Also on the theoretical side the treatment of larger and larger complex systems still poses significant challenges. However the development of time dependent DFT with coevolution of quantum optical fields is a promising route for studying ultrafast quantum optical phenomena on the nanoscale~\cite{ruggenthaler18}.

\subsection{Coherent spectroscopy on the nanoscale}
\label{sec:specnano}

\begin{figure}[h!]
\center\includegraphics[width=\columnwidth]{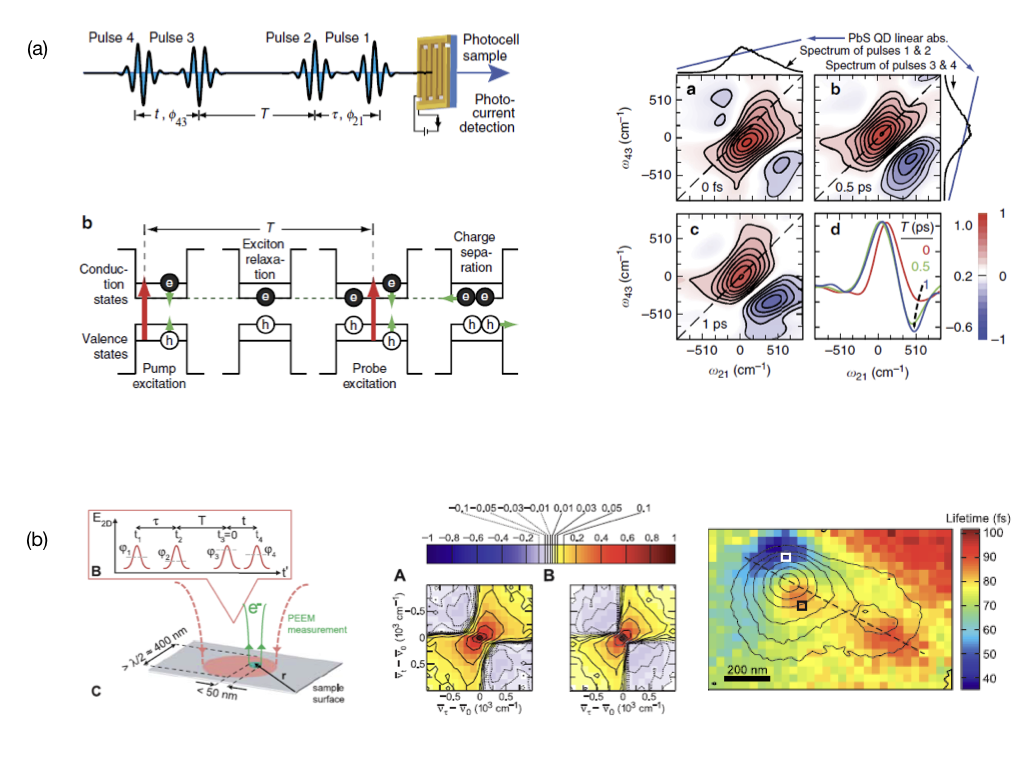}
\caption{Coherent two-dimensional spectroscopy of nanoscale systems. (a) 2D photocurrent spectroscopy of a PbS QD photocell. The left part shows the interaction of the four different pulses used for excitation of the photocell and the excitation scheme for exciton and multi-exciton generation in upper and lower part, respectively. The right part shows the 2D spectra at high excitation density for different delays T revealing  multi-exciton formation influencing the photocurrent signal (reprinted Fig. 1, 3 and 4 with permission from Ref.~\cite{karki14}. Copyright (2014) by the Springer Nature Publishing Group). (b) Coherent  2D nanoscopy based on photoelectron detection using a PEEM. Here the final state population excited by a four pulse sequence is a photoemission state. Coherent 2D spectra detected for different pixels of the emission pattern yield various coherence life times of plasmonic excitations on a corrugated Ag film. In the right part  the a so measured lifetime map is shown. (reprinted Fig. 2 and Fig. 3 with permission from Ref.~\cite{aeschlimann11}. Copyright (2011) by the American Association for the Advancement of Science).\label{fig:specnano}}
\end{figure}

Coherences and correlations are essential features of quantum phenomena. The coherent superposition of wavefunctions determines quantum states and all macroscopic observables. Revealing the nature of coherences and correlations in nanoscale quantum systems is crucial for a better understanding of quantum effects and their connection to our macroscopic world. Apart from this fundamental significance, the frontier between coherent and incoherent phenomena has many practical consequences for the understanding, design, and function of natural or artificial (nanoscale) objects. For example, in the context of energy transfer one can ask over which time and length scales transport occurs coherently or incoherently and what can be done to control the boundary between these regimes. Advanced and newly developed methods of spectroscopy play an important role for investigating coherent dynamics in nanoscale systems.
Conventional coherent spectroscopy, especially time-resolved multidimensional coherent spectroscopy, already provides detailed information about nanoscale and even intramolecular energy transfer and coupling phenomena~\cite{brixner05c,falke14,haedler15,desio16}. Examples for research on coherent nanoscale dynamics include the nonlinear spectroscopy of coherences in semiconductors~\cite{karki14} (Fig.~\ref{fig:specnano}a), identifying quantum coherent dynamics. In nanophotonics, these concepts need to be expanded to heterogeneous systems, in which a finite number of individual quantum emitters must be considered. On the nanoscale, the discreteness of quantum emitters is important since it affects the dynamics and the collective behavior.  Presently rather simple coupled systems are considered~\cite{otten15} showing already rich dynamics and the possibility to manipulate quantum coherences using robust coherent control schemes~\cite{rousseaux16}.
NQO benefits substantially from coherent spectroscopy with nanoscale spatial resolution, i.e. beyond the diffraction limit~\cite{aeschlimann11}. The probed linear and nonlinear responses contain more information than the integral responses measured in the far field. In addition, inhomogeneous line-broadening that might mask coherent dynamics in measurements performed on an ensemble is absent. Combination of methods adapted from conventional coherent spectroscopies with optical near-field probes (scanning near-field optical microscopy [SNOM], nanotips)~\cite{bechtel14,kratsov16} or photoemission electron microscopy (PEEM)~\cite{aeschlimann11} has significantly expanded the methodologies to probe nanoscale coherent dynamics~\cite{collini13}. The first implementation of a PEEM based coherent spectroscopy with nanometer spatial resolution~\cite{aeschlimann11} is shown in Fig.~\ref{fig:specnano}b. The effect of the multi-pulse excitation employed in coherent spectroscopy is read out here by detecting photoelectrons which can be achieved in commercial PEEM setups with down to a few nm spatial resolution. Based on such techniques the complex dynamics in individual quantum optical devices will become accessible for detailed investigations. The local nonlinear response contains information about coherent energy transfer in nanoscale systems. Combining multidimensional spectroscopy methods with nanoscale resolution avoids the effects of system inhomogeneity and yields access to local response functions. This entails combining various aspects of advanced spectroscopy, nano-optics and quantum optics, such as nanoscale energy transport, quantum optical coupling mechanisms, light confinement and trapping.

A broad range of nanophotonic systems is investigated using coherent spectroscopies. Both far-field methods and techniques that allow for nanoscale resolution are employed. Examples include: exciton-plasmon coupling, and disordered photonic materials. Spatiotemporal optical near- field control in combination with coherent spectroscopy allows probing the nonlocal response in complex nanostructures. Nonlocal coupling phenomena reaching over distances more than a few nanometers become accessible. Improved spatial resolution will provide access to coherent dynamics at the single-molecule level. This requires, among others further enhanced spatiotemporally resolved coherent nanospectroscopy with sub-10 nm resolution and single-molecule sensitivity.

\subsection{Active nanophotonics}
\label{sec:actnano}

\begin{figure}[h!]
\center\includegraphics[width=\columnwidth]{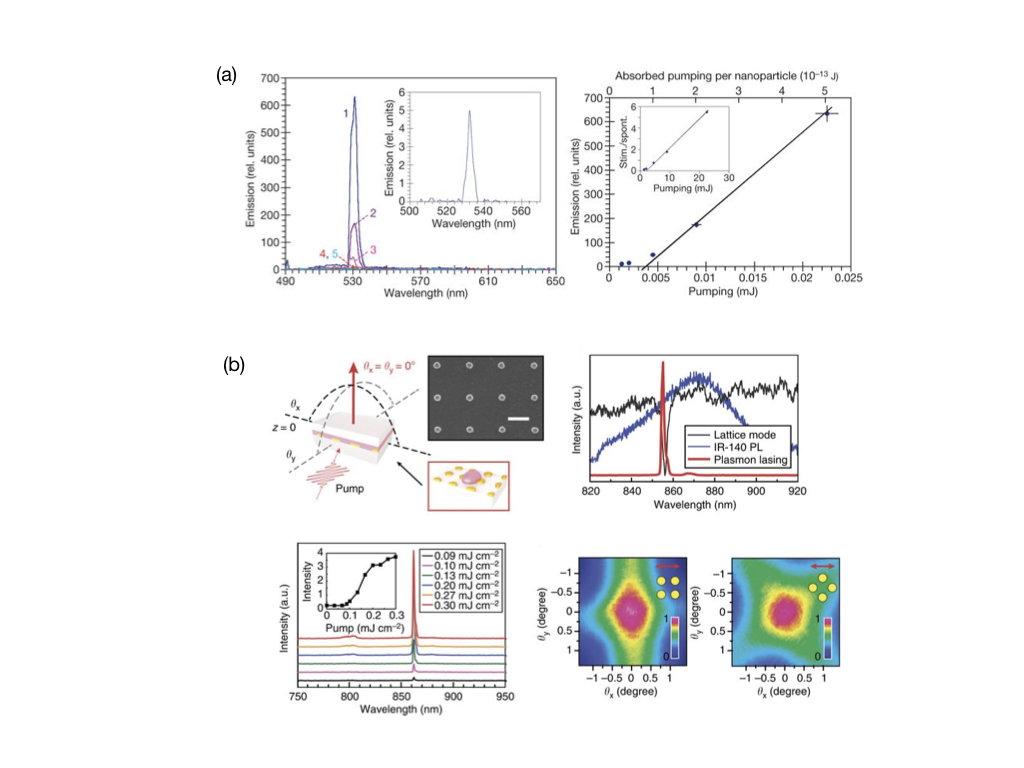}
\caption{(a) The hybrid nanoparticle architecture, indicating dye molecules throughout the silica shell and stimulated emission spectra of the nanoparticle sample pumped with 22.5 mJ (1), 9 mJ (2), 4.5 mJ (3), 2 mJ (4) and 1.25 mJ (5) 5-ns optical parametric oscillator pulses at $\lambda$ = 488 nm (reprinted Fig. 4 with permission from Ref.~\cite{noginov09}. Copyright (2009) by the Springer Nature Publishing Group). (b) Lasing emission from Au NP arrays on transparent substrates sandwiched between IR-140 dissolved in an organic solvent and a glass coverslip (reprinted Fig. 1 with permission from Ref.~\cite{yang15}. Copyright (2015) CC License 4.0).\label{fig:actnano}}
\end{figure}

Nanoplasmonic resonators embedded in an optically active medium allow the fabrication of nanoscale laser sources with subwavelength dimensions known as spasers~\cite{bergman03,noginov09,premaratne17} (see Fig.~\ref{fig:actnano}a). The localized plasmon acts as a low-Q cavity and the photons in this cavity mode are amplified using energy provided by the pumped, active medium and radiated to the far field via the radiative damping channel of the localized plasmon. Plasmon lasers can support ultrasmall mode confinement and ultrafast dynamics with device feature sizes below the diffraction limit. 
The demonstration of efficient sub-wavelength coherent light sources and amplifiers attracted great interest. However, the implementation of spasers in photonic nanocircuits requires electrical pumping schemes~\cite{fedyanin12}. Beyond the spaser based on a single nanoantenna surrounded by an optical gain medium interesting applications arise from combining metamaterials with optical gain~\cite{hess13}.
Emission of transverse radiation relies on mode amplification of a lossy cavity, i.e. an electromagnetic mode that couples to the far field. It is also important to study optical amplification beyond the effective medium gain model and master equations including many body effects of individual emitters to understand the spasers and nanoscale plasmonic resonators. 
The first demonstration of spasers relies on dye molecules embedded in a polymer shell surrounding a gold nanoparticle~\cite{noginov09}. Optical pumping leads to fluorescence that narrows at some pumping level indicating amplification of photons in the plasmon mode of the nanoparticle. Later, nanowire lasers and plasmonic nanoantannes as resonators have been used to obtain lasing.  
Most plasmon-based nanolasers rely on solid gain materials such as inorganic semiconducting nanowire or organic dye in a solid matrix. Plasmon nanolaser whose emission properties can be modulated in real time is based on liquid gain dyes. This gain medium with different refractive indices were able to tune the lattice plasmon resonance, and hence lasing wavelength, over the entire bandwidth of the dye~\cite{yang15}. Furthermore, single QD lasers and thresholdless lasing with a broadband gain medium are obtained~\cite{khajavikhan12}. Recently, lasing both in dark and bright modes of an array of silver nanoparticles combined with optically pumped dye molecule have been also obtained~\cite{hakala17}.

Although there has been promising progress, the small mode volume of plasmonic nanoresonators requires high population inversion densities to allow nanolasing. In addition, for future applications electrical pumping schemes and non-radiative amplifiers are required.
The generation of coherent radiation by these plasmonic resonant structures has attracted considerable interest due to its potential in applications ranging from on-chip optical communication to ultrahigh-resolution and high-throughput imaging, sensing and spectroscopy. Active nanophotonics research is aimed at developing the ‘ultimate’ nanolaser which is efficient, has a low lasing threshold, operates at room temperature and occupies a small volume on a chip.

\subsection{Coherent control at the nanoscale}
\label{sec:ccnano}

\begin{figure}[h!]
\center\includegraphics[width=\columnwidth]{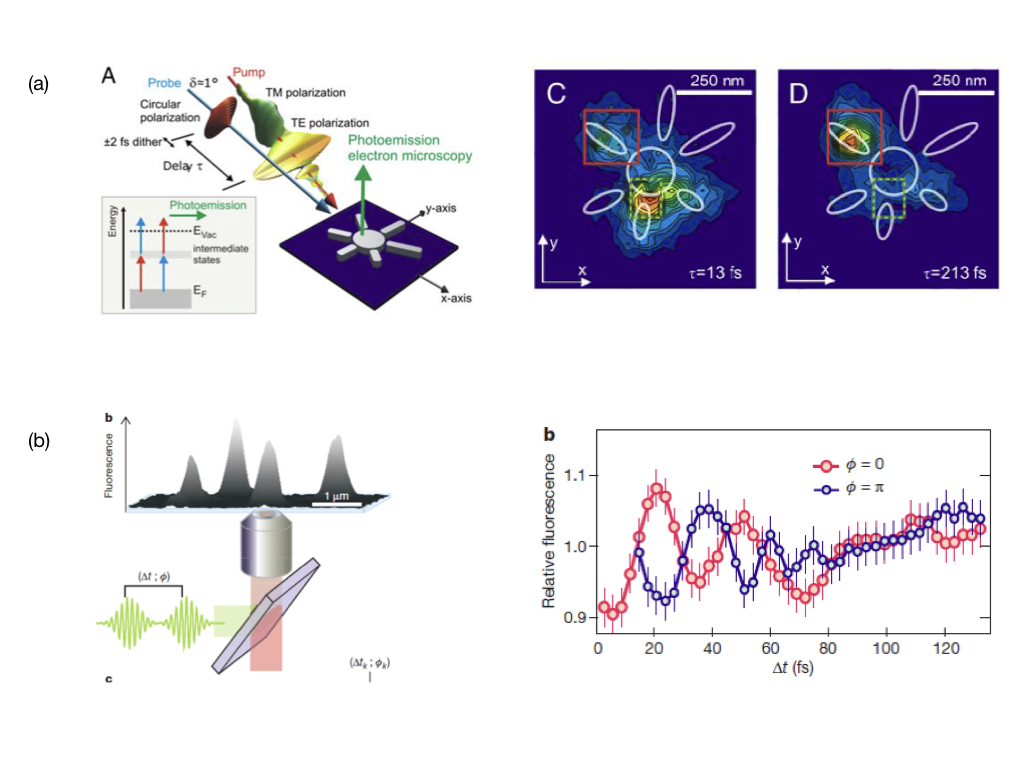}
\caption{Coherent control below the diffraction limit. (a) Demonstration of spatio-temporal excitation control in the vicinity of an optical nanoantenna via polarization pulse shaping. The left panel depicts the experimental implementation. The local excitation generated by a polarization-shaped pulse in the nanostructure is probed by a second time-correlated pulse. The right panel shows the local excitation pattern for two different times. (reprinted Fig. 2 and Fig. 3 with permission from Ref.~\cite{aeschlimann10}. Copyright (2010) by the National Academy of Sciences). (b) Coherent control of vibrational wave packets in single molecules. Pulse shaping technology is combined with high numerical aperture illumination as it is used in single-molecule spectroscopy. Via the phase and delay between the excitation pulses the molecular vibration of a single molecule is controlled and detected via the fluorescence signal. (reprinted Fig. 1 and Fig. 3 with permission from Ref.~\cite{brinks10}. Copyright (2010) by the Springer Nature Publishing Group).\label{fig:ccnano}}
\end{figure}

In wave mechanics the phase of excitation and driving fields is essential and offers the opportunity for control. This holds for classical wave phenomena and quantum states in a rather similar fashion. Many concepts were first demonstrated in nuclear spin manipulation schemes developed in the field of nuclear magnetic resonance spectroscopy. With the development of ultrafast laser technology means to control quantum systems by light became available with molecules acting as a second testbed for demonstrations of coherent control schemes. Besides control of chemical reactions and nuclear spins, the field has in recent years significantly expanded to other realms. Particularly relevant in the context of future QT are applications of coherent control schemes in nanophotonics~\cite{aeschlimann07,brinks10,hildner13}. They allow for addressing and manipulating individual quantum systems and provide new insights about quantum coherences that are then no longer hidden in an inhomogeneous ensemble. In nanophotonics a classical coherent control scheme has been demonstrated to enable spatio-temporal control of nanophotonic excitations (Fig.~\ref{fig:ccnano}a). Tailoring the polarization pulse shape of incident light fields allows controlling the spatial and temporal evolution of nanoscale excitations in the vicinity of suitably designed nanoantenna structures~\cite{aeschlimann10}. This opens a pathway to control the excitation of coupled quantum systems in an unprecedented fashion enabling possibly new quantum functionalities. For example it is conceivable that a spatio-temporally controlled excitation of a complex multipartite quantum systems leads to new correlation effects for emitted fluorescence photons. It is anticipated that the development of multidimensional coherent spectroscopy with nanometer spatial resolution~\cite{aeschlimann11}, together with spatially selective addressing in optical near- fields, enables complex quantum protocols in individual nanoscale assemblies of interacting qubits. The rapid advances in nanotechnology complement this development. Precise fabrication of high quality nanostructures with deterministic coupling to single qubits allows investigating complex coherent phenomena. In the long-term, this field offers a promising route towards an alternative system platform for highly integrated QT.

The feasibility of single system quantum coherent control schemes was demonstrated impressively by controlling vibrational wave packets in single molecules (Fig.~\ref{fig:ccnano}b)~\cite{brinks10}. This demonstrates that quantum dynamics in single complex nanoscale quantum devices can be controlled in a most flexible way. The coherent control schemes developed in chemistry need adaptation for applications in nano-optical settings, where longitudinal fields become relevant, expanding the degrees of freedom for control. Theoretical efforts concern, for example optical near- field control, novel quantum control schemes adapted to nanoscale systems, controlling nanooptical photon-photon interactions, and exploration of adaptive versus open-loop control.
Nanoantennas serve as gateway to selective nanoscale quantum systems such as chromophores or QDs acting as qubits. The anisotropic response of such antennas enables selective spatiotemporal excitation schemes, based, for instance, in polarization pulse shaping assisted optical near- field control, time-resolved single- molecule spectroscopy, quantum emitters coupled to nanoantennas, hybridization or strong coupling between quantum emitters and nanoantennas, and 2D materials and their heterostructures. Presently the nanoantennas used to selectively address individual quantum systems are conceived as classical entities. With increasing coupling strength, this picture fails and a fully quantum description of the combined system interacting with the environment is needed. In nanoscale quantum devices the effects of decoherence and dissipation must be minimized by developing robust control schemes~\cite{rousseaux16} addressing minimization of dissipation and decoherence, realistic quantum description of system-bath interaction for complex nanodevices, and development of robust coherent control schemes. 

\section{Nanoscale quantum coherence}

\subsection{Quantum coherence and dephasing as a sensing tool}

\begin{figure}[h!]
\center\includegraphics[width=\columnwidth]{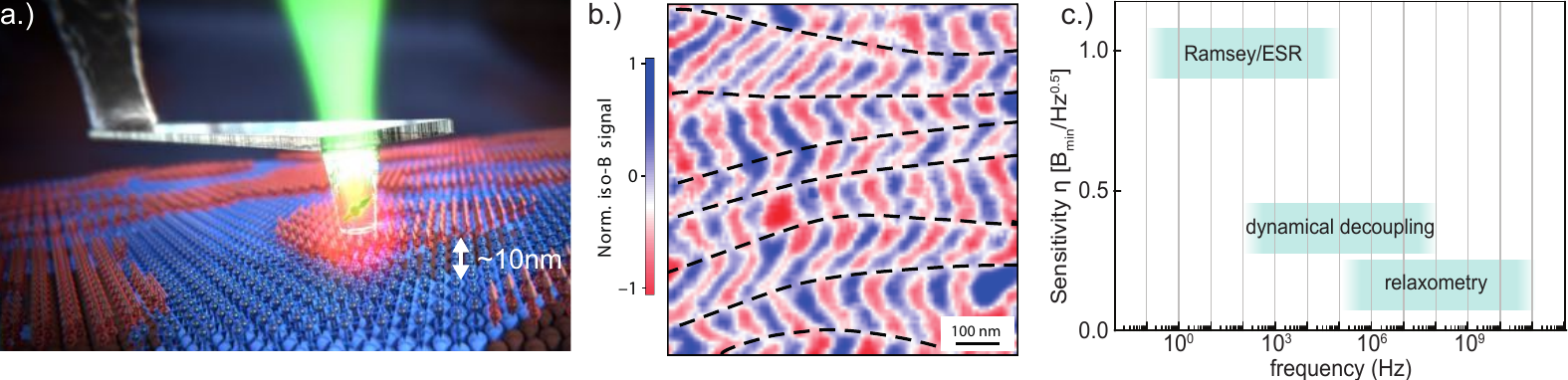}
\caption{Quantum sensing and magnetic imaging with single spins. a) Artist impression of a single-spin microscope used for high-sensitivity, nanoscale imaging of, e.g., non-trivial magnetic textures. The structure hosting the sensor spin (green) is a nanophotonic device in form of a nanopillar, which enhances the efficiency of optical addressing of the sensor spin. b) Example of a nanoscale magnetic image obtained by scanning NV magnetometry. The image shows the stray magnetic field generated by a spin cycloid in the compound BiFeO$_3$, which was observed for the first time in the work of Ref.~\cite{gross17} (reprinted Fig.~2c with permission from Ref.~\cite{gross17}. Copyright (2017) by the Springer Nature Publishing Group). c) Overview of common quantum sensing modalities with single spins. The sensing modalities differ in the coherent manipulation sequences employed and lead to different sensitivities and sensing bandwidths, which can be readily adapted to the sensing task at hand.\label{fig:NVMag}}
\end{figure}

Quantum systems are highly susceptible to external stimuli such as magnetic or electric fields, temperature changes or strain~\cite{lee17,teissier14,barfuss18,barson17}. Consequently, they are amenable to precisely sensing these quantities, often with nanoscale spatial resolution, by virtue of the inherently miniature size that many solid-state quantum systems offer~\cite{balasubramanian08,kraus14}. The sensitivity at which such measurements can be performed is limited by the quantum sensor's coherence time~\cite{rondin14,degen17}, i.e. the timescale over which it preserves its quantum phase. Specifically, for a quantum projection noise limited measurement of magnetic fields using a single spin, as schematically depicted in Fig.~\ref{fig:NVMag}a, the magnetic field sensitivity is fundamentally set by\,\cite{budker07}
\begin{equation}
\eta=\frac{1}{\gamma}\frac{1}{\sqrt{T_{\rm coh}}}
\label{EqSens}
\end{equation}
where $\gamma$ is the gyromagnetic ratio of the spin and $T_{\rm coh}$ the coherence time relevant for a given sensing sequence. In presence of classical readout noise, the sensitivity is further worsened by a factor of $C$~\cite{taylor08,degen17}, characteristic of spin readout contrast and efficiency, so that $\eta=\frac{1}{C \gamma}\frac{1}{\sqrt{T_{\rm coh}}}$. One typically finds $C\sim10^{-2}-10^{-3}$, e.g. for single NV center spins in diamond operated under ambient conditions, which for continuous-wave sensing (where $T_{\rm coh}=T_2^*\sim1~\mu$s is the inhomogeneous dephasing time) yields $\eta\sim1-10~\mu$T/Hz$^{0.5}$. 
These sensitivities can be further improved by optimising any of the free parameters occurring in Eq.\,\ref{EqSens}. Specifically, nanophotonics and quantum logic can be employed to improve the spin readout efficiency and thus parameter $C$. Similarly, quantum control and dynamical decoupling~\cite{taylor08,degen17} offer ways to protect coherence, prolong $T_{\rm coh}$ and thereby improving $\eta$. Both boost sensitivities significantly and in combination can yield, for example, quantum magnetometers that reach single-protein sensitivity~\cite{lovchinsky16}.

While internal decoherence limits a quantum sensor's performance~\cite{degen17}, decoherence caused by external sources, i.e. additional environmental fluctuations or entanglement with the environment, offers a valuable resource for sensing. Indeed, a quantum sensor in proximity to a sample may experience excess dephasing, which may carry relevant information. Such decoherence microscopy has been proposed in the context of life sciences~\cite{hall10} and fundamental solid- state physics~\cite{vannieuwenburg16}. One key strength of decoherence-based sensing is that it does not require phase-stable signal sources and that quantum control can be used to tailor the spectral response of the quantum sensor to either lock the sensor to a particular sensing frequency or perform noise spectroscopy of the environment~\cite{degen17}. 

Initial experimental results have validated the potential of quantum decoherence microscopy and demonstrated it's applicability to nanoscale sensing~\cite{schmid-lorch15,luan15}. Promoting these concepts to real-life applications in life sciences~\cite{hall10} (where for instance ion channels in cell membranes could be investigated) or hard-to-address open problems in solid-state physics~\cite{vannieuwenburg16,rodriguez18} (such as studying strongly correlated electron systems) remains an open challenge. However, recent advances in quantum sensing technologies, such as robust, coherent scanning quantum systems~\cite{maletinsky12,appel16} that employ innovative nanophotonic concepts~\cite{neu14}, or the demonstration of nanoscale quantum sensing at cryogenic temperatures~\cite{thiel16}, render this a highly promising avenue in quantum engineering research. 

The fact that startup companies~\cite{qnami18} have emerged that aim at marketing such technologies to a broader range of customers is a further indication of the attractiveness and broad applicability of coherence- or decoherence- based quantum sensing technologies. However, challenges related to reproducibility and robustness have arisen and include in particular, the availability of highly coherent quantum probes at the nanoscale, the development of highly efficient sensing protocols based on decoherence and, ultimately, Identifying and demonstrating key applications outside the field of quantum sciences.

\subsection{Coherent quantum transport for energy harvesting}

Nanoscale transport/bio-inspired materials such as light-harvesting complexes could open new routes towards efficient photovoltaic cells. There is theoretical and experimental evidence that coherent (quantum) transport is crucial for ultrafast energy transfer, which is in turn core to the efficiency of the photosynthesis processes of energy harvesting and energy transport~\cite{bredas16,wientjes14,panitchayangkoon10}. A key role in this process is also played by the structural order/disorder interplay, which is conducive to a regime where population can be transferred between semilocalized and extended states, favouring coherent transport.

Coherence/decoherence interplay is fundamental for a step-change increase of excitation diffusion lengths, another important factor for energy harvesting and transport efficiency. Here coherence provides the memory necessary to ensure a degree of spatial directionality in an otherwise inefficient random walk of excitations~\cite{bredas16}. The interplay between timescales, the importance of (quantum) correlations, the role of disorder, and the different coupling regimes between the components of these complex systems require the development of more accurate models to describe the natural energy-harvesting and transport processes, including going beyond F\"orster theory and Born-Oppenheimer approximation~\cite{bredas16,turner17}.

Light-harvesting complexes~\cite{wientjes14,panitchayangkoon10,chen15} containing pigments and proteins are used by plants and photosynthetic bacteria to efficiently harvest solar energy. In recent years, synthetic light- harvesting arrays, e.g. using boron dipyrromethene dyes and pyrene~\cite{ziessel13} or artificially enhanced natural biological systems~\cite{yoneda15}, have also been developed. Currently biological systems are the only known very efficient light harvesters. This is due to the fact that the most efficient methods have been selected through evolution. Identifying and reproducing these mechanisms in photovoltaic devices may be crucial to reach a step-change in their performance and to increase efficiency in artificial  photosynthetic cells. The main challenges and goals for the current technology are to develop new materials that have to reduce excitonic recombination that leads into an increase of diffusion lengths. In the end, the final aim of current research in this field is to develop self-organized, and self-sustained artificial photosynthetic cells for future applications.
 
\subsection{Fundamental aspects of quantum coherence at the nanoscale}

Quantum coherence has been demonstrated in various interference experiments with electrons, neutrons, atoms, molecules and more recently aggregates counting up to 10000 amu~\cite{gerlich11}. It is extremely difficult to attain quantum interferences (Schr\"odinger-cat experiments) with larger objects~\cite{arndt14} because (i) the de Broglie wavelength is inversely proportional to the mass of the object, while (ii) environmental decoherence increases exponentially with its size, which drastically reduces the coherence time during which it is possible to observe interferences. During the last decade, various mesoscopic systems (from 1 billion amu and beyond) have been conceived, developed and studied in the laboratory in extreme vacuum conditions and at extremely low temperatures~\cite{yuan15,ares16}, aiming at implementing Schr\"odinger-cat experiments, and there is widespread hope in the community that it will be possible soon to operate them in the quantum regime.

In order to investigate the mesoscopic transition it is imperative to take into account quantum decoherence and from a fundamental point of view to develop nonstandard models (gravitationally induced localization and/or decoherence~\cite{kafri14,scala13,yang13}, spontaneous localization~\cite{vinante16}), which predict a quantum to classical transition in this regime. Essentially all planned experiments in this field require OM devices at some level, but otherwise versatility is the rule. Therefore, the broad range of nanoscale systems, materials and techniques are relevant for exploring the fundamental aspects of quantum coherence
Furthermore, realizing quantum interferences in the mesoscopic regime would not only open new perspectives in sensing and metrology, but also open the door to fundamental experimental tests related to the nature of gravity~\cite{kafri14,scala13,yang13} at the quantum level and/or to the different quantum measurement issues~\cite{vinante16,kolaric18}.

\subsection{Interaction of entangled light with nanostructures}

\begin{figure}[h!]
\center\includegraphics[width=0.8\columnwidth]{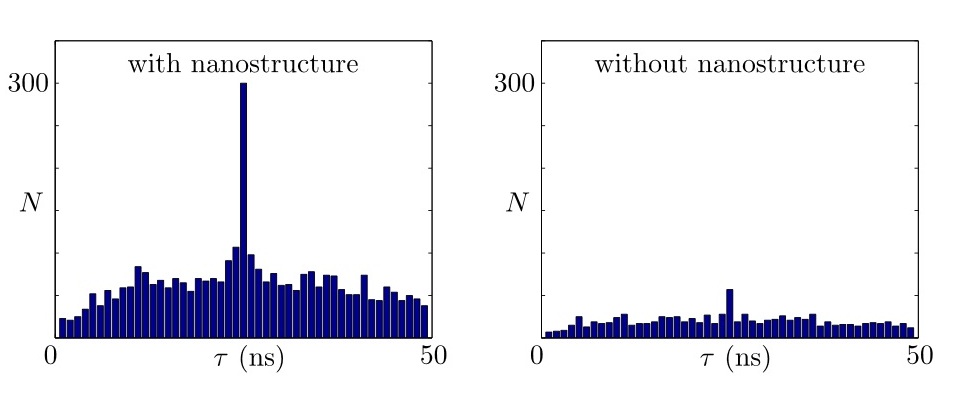}
\caption{Coincidence peaks. Coincidence peaks obtained after one photon of the entangled pair is transferred through either the plasmonic structure at normal incidence (left) or a flat gold film (right) (reprinted Fig. 4 with permission from Ref.~\cite{olislager15}. Copyright (2015) by Walter de Gruyter GmbH).\label{fig:FigE}}
\end{figure}

A consequence of coherence in nanoscale objects is the possibility of achieving entanglement within such systems. Entanglement is a key property by which quantum light differs from classical light. It has been shown in the past that the polarisation and the energy degrees of freedom of entangled photons could be successfully transferred to collective plasmonic modes, without degrading the quality of the entanglement. However, directly generating entanglement in a controlled way would often require being able to fully control the interaction between such systems. Different model experiments have studied the interaction of entangled photons with surface plasmons, demonstrating that the process of light conversion into plasmons and then back to photons preserves entanglement, both for polarization~\cite{altewischer02} and for time-energy~\cite{fasel05,olislager15} entangled photons. Interaction of frequency-bin entangled photons  with a nanostructure is presented in Fig.~\ref{fig:FigE}.

Figure~\ref{fig:FigE} shows typical results regarding interaction of entangled light with nanostructures and confirms that entanglement is preserved due to presence of plasmons that facilitate photon-plasmon-photon conversion. Present experiments can be the basis for further studies on the link between the nature of plasmonic resonances and the photon-plasmon-photon conversion processes. It may also contribute to the development of applications of metallic nanostructures in quantum-based technologies. Additionally, it would be interesting to harness entanglement for various spectroscopic and chemical studies.

Two-photon interactions with entangled light would also be relevant, for instance to probe a system. Recently, several schemes have been proposed demonstrating the advantage of performing spectroscopy beyond the classical paradigms by making use of such entangled states of light~\cite{dorfman16}. However, no experimental implementations of those theoretical ideas with energy-entangled photons have been demonstrated up to now. The biggest challenges and goals regarding applications of entangled light include the development of the new spectroscopy schemes using entangled light and possibility to transfer entanglement from light to large collective structures and atomic systems for future quantum memories and quantum repeaters.

\subsection{Strong light-matter interaction at ambient conditions}

\begin{figure}[h!]
\centering
\includegraphics[width=0.5\columnwidth]{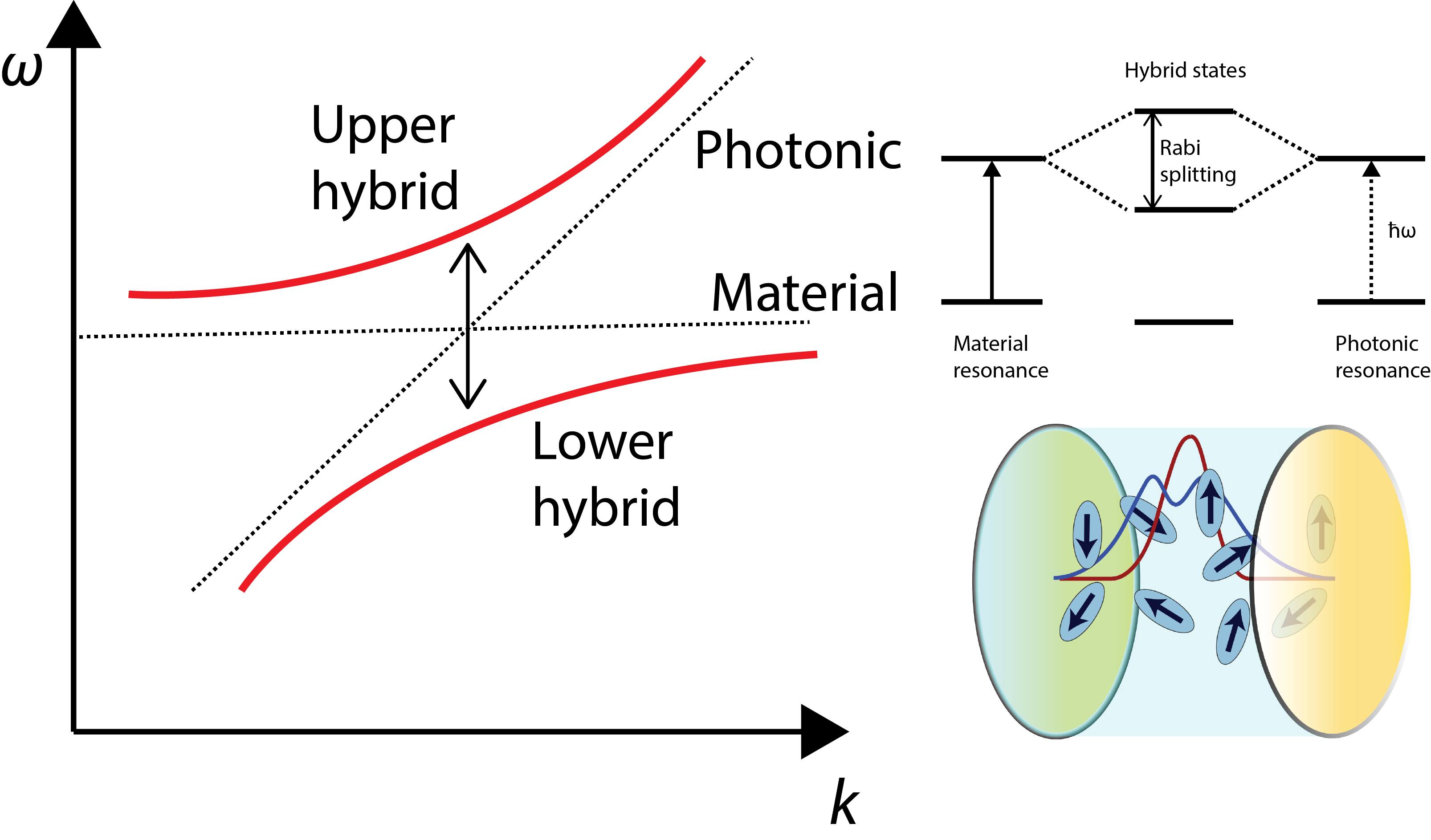}
\caption{The strong interaction between the material entity and light within plasmonic/photonic cavities creates new hybrid light–matter states that have significantly different energy levels from those of the material entity and of the optical system individually (adapted Fig.~1 with permission from Ref.~\cite{kolaric18}. Copyright (2018) by Wiley-VCH Verlag GmbH \& Co. KGaA).\label{fig:strongcoupling}}
\end{figure}

The interaction between a quantum emitter and its local electromagnetic environment is typically weak, such that only the spontaneous decay rate is modified and the emission frequency remains unaltered~\cite{torma15}. Strangely enough, in the case of strong coupling it is conceivable for light and molecules to bond, creating new hybrid light-matter states, i.e., dressed states or polaritons ~\cite{torma15,ebbesen16, kolaric18} with far-reaching consequences for these strongly coupled materials. Schematic view of strong coupling is presented in Figure~\ref{fig:strongcoupling}.

Even stranger, there is no `real' light needed to obtain the effects, it simply appears from the vacuum, creating `something from nothing'. Surprisingly, the setup required to create these materials has become moderately straightforward. In its simplest form, one only needs to put a strongly absorbing material at the appropriate place between two mirrors or within open plasmonic cavity  and quantum magic can appear. Only recently has it been discovered that strong coupling can affect a host of significant effects at a material and molecular level, which were thought to be independent of the `light' environment: phase transitions, conductivity, chemical reactions, etc.\cite{kolaric18,wang14,orgiu15,thomas16}.

Up to now, different photonic, plasmonic or hybrid plasmonic-photonic structures~\cite{ebbesen16,kolaric18} with unique resonance patterns are used to create an environment in which strong coupling occurs. Developing this unique photonic/plasmonic environment that could effectively control and mould strong light-matter interactions at room temperature is the primary goal that could revolutionize current technology and significantly reshape modern chemical science.

Strong coupling profoundly connects material science with fundamental physics and quantum information science, and offers the possibility to control and mould material and molecular properties. The general application of this concept could lead to a new scientific revolution, and could be applied to engineer novel complex materials, such as high-temperature superconductors, topological materials, semiconductors, etc. It can be employed to control biological dynamics and to affect chemical reactivity in an unprecedented way, just by using and harvesting interactions between the vacuum field and material oscillators. Additionally, strong coupling offers a quantum mechanical control on classical processes, such as phase transitions and self-assembly, creating a remarkable link between the quantum and classical realm. Furthermore, engineering structures for strong coupling could rapidly lead to new breakthroughs in the field of nanoscale optics and photonics.  Regardless of the fact that, at the moment, strong coupling is still at its infant stage of development, its already attract the interest of the broad material science community and could open new horizons in chemical, physical and material research and technology.

\section{Cooperative effects, correlations and many-body physics tailored by strongly confined optical fields}

The tight confinement of light in nanophotonic structures drastically increases optical nonlinearities as well as the coupling between photons and individual quantum emitters such as QDs, defect centers or atoms trapped nearby. Under such  conditions, photons can efficiently mediate long-range interactions between matter excitations or even behave themselves as strongly interacting particles. The localization of optical fields to a small region in space enhances as well the optomechanical coupling to isolated vibrational modes, which can lead to novel types of quantum nonlinear effects at the level of single photons and phonons.  By tailoring these different forms of light-matter interactions in cavities, waveguides and photonic lattices, a plethora of possibilities for studying complex quantum optical phenomena arise. This includes cooperative and ultrastrong coupling effects, non-equilibrium phase transitions, and the simulation of other strongly-correlated many-body systems using light. 

\subsection{Atom-light interactions in one dimension}

\begin{figure}[h!]
\center\includegraphics[width=0.7\columnwidth]{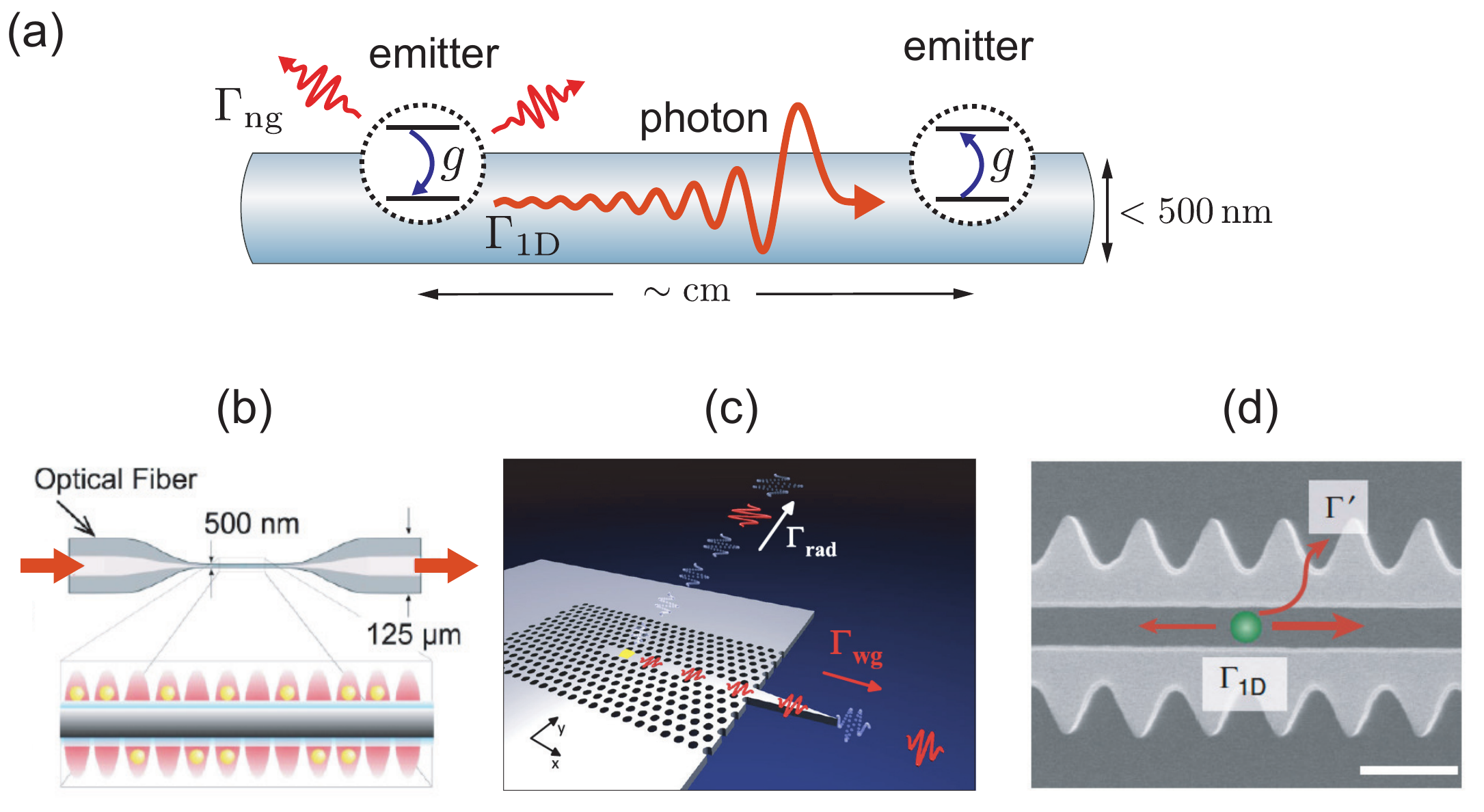}
\caption{(a)~Sketch of a waveguide QED setup with two quantum emitters coupled to a 1D photonic channel. If the rate of emission into the waveguide, $\Gamma_{\rm 1D}$, exceeds the emission rate $\Gamma_{\rm ng}$ into other, non-guided modes, efficient long-range interactions between multiple emitters can be attained. This 1D regime of light-matter interactions can be implemented, for example, with  (b) trapped atoms coupled to the evanescent field of a tapered nanofiber (reprinted Fig.~1 with permission from Ref.~\cite{dawkins11}. Copyright (2011) by the American Physical Society), (c) QDs coupled to a photonic-crystal waveguide (reprinted Fig.~1a  with permission from Ref.~\cite{arcari14}. Copyright (2014) by the American Physical Society), or (d) by combining cold atoms and nanophotonic crystal structures in a so-called alligator waveguide setup (reprinted Fig.~1a with permission from Ref.~\cite{goban14}. Copyright (2014) by the Springer Nature Publishing Group).\label{fig:WaveguideQED}}
\end{figure}

In nanophotonic fibers and waveguides, light is transversely confined to an area less than a wavelength squared, which is smaller than the resonant scattering cross section of a single emitter. At the same time, the photons can propagate freely along the waveguides over long distances. This combination makes such \emph{waveguide QED} settings particularly interesting for quantum networking applications as well as for the study of exotic quantum many-body systems with `infinite-range' interactions. To reach this one dimensional regime in an actual experiment, the rate $\Gamma_{\rm 1D}$ at which an excited emitter decays into the waveguide must exceed the decay rate $\Gamma_{\rm ng}$ into all non-guided modes as sketched in Fig.~\ref{fig:WaveguideQED}a. This condition is usually expressed in terms of the $\beta$-factor
\begin{equation}
\beta= \frac{\Gamma_{\rm 1D}}{\Gamma_{\rm 1D}+\Gamma_{\rm ng}} \leq 1.
\end{equation}
Experimental values for this parameter range from $\beta\sim 0.1$ for clouds of atoms that are coupled evanescently to tapered nanofibers~\cite{dawkins11} (see Fig.~\ref{fig:WaveguideQED}b), up to values of $\beta\sim 0.99$ for single QDs inside photonic-crystal waveguides~\cite{arcari14} (see Fig.~\ref{fig:WaveguideQED}c). To explore the full range of waveguide QED physics, a current challenge is to combine large $\beta$-factors with a large number of nearly identical emitters. The trapping of cold atoms in close vicinity of optimized nanophotonic structures~\cite{goban14} (see Fig.~\ref{fig:WaveguideQED}d), or the use of individually tunable QDs are two potential approaches to reach this goal. 

The 1D confinement of light and the efficient coupling of photons to individual emitters gives  rise  to  many  intriguing  phenomena  and  applications,  such  as  single  photon  switches  and  mirrors~\cite{shen05}, the generation of long-range entanglement~\cite{stannigel12}, or self-organized atomic lattices~\cite{chang13}. In nanophotonic platforms, the exploitation of near-field effects can further be used to implement strong chiral light-matter interactions, in which case the emitters only interact with photons propagating along a single direction~\cite{lodahl17}. This unusual effect provides a crucial ingredient for quantum communication protocols between multiple emitters, where otherwise in each emission event photons would be emitted to the left and the right and would be lost with 50\% probability. 

The physics of light-matter interactions in 1D becomes even more involved in the presence of band edges and  band gaps,  near which  the  group  velocity of photons is strongly reduced or free propagation is completely prohibited. Under such conditions the usual Markovian description of radiative processes breaks down and so-called atom-photon bound states emerge~\cite{lambropoulos00}. These states represent long-lived excitations with an energy outside the propagation band, where the excited state of the emitter is dressed by an exponentially localized photon. Therefore, this setting interpolates between a more cavity-like and a more waveguide-like scenario, where the range of photon-mediated interactions  is determined by the extent of the photonic component of the bound state.  By controlling this length scale via the emitter-photon detuning, unprecedented possibilities for engineering long-range emitter-emitter interactions and effective spin models with fully controllable couplings~\cite{chang18} emerge. 

\subsection{Collective effects and phase transitions}

\begin{figure}[h!]
\center\includegraphics[width=0.7\columnwidth]{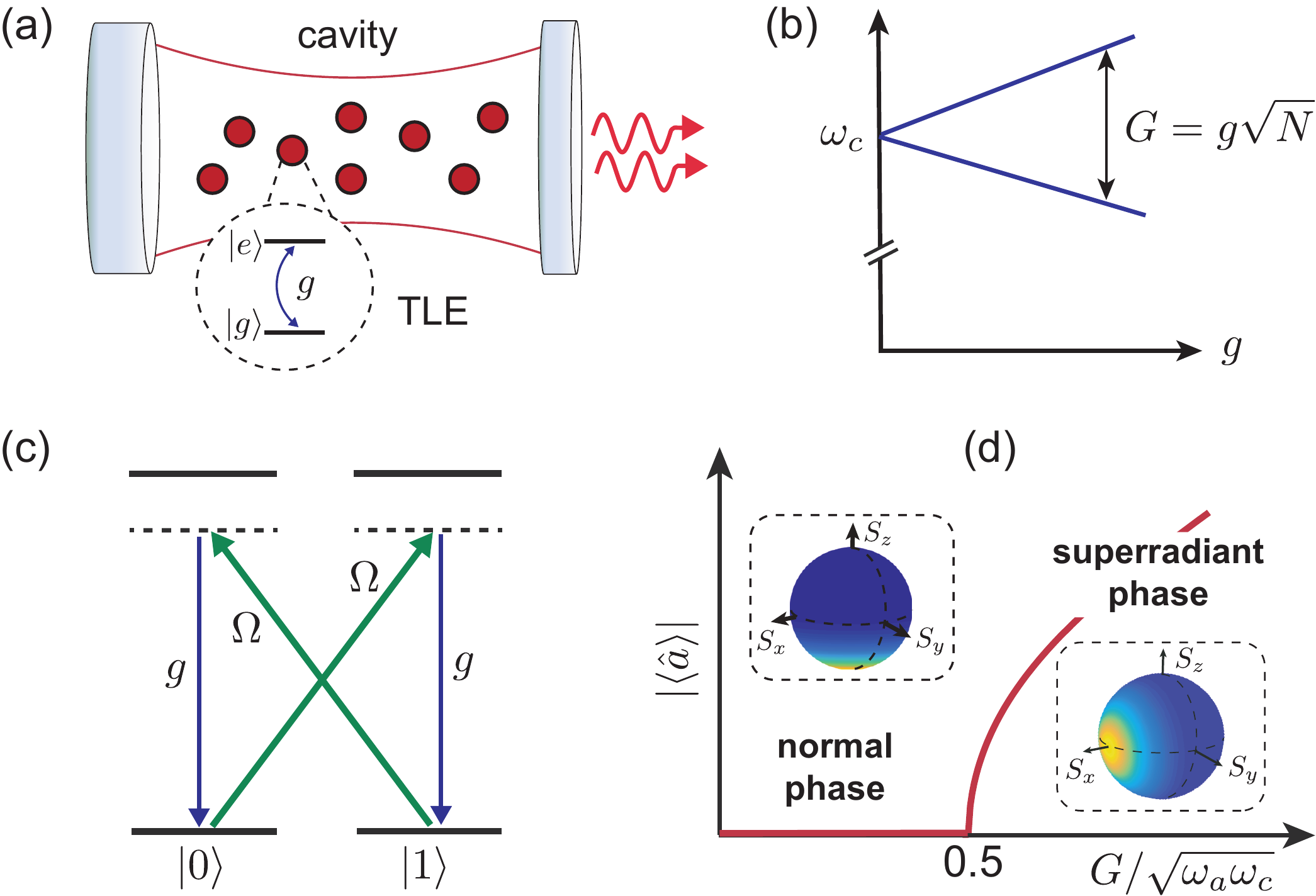}
\caption{(a) Sketch of a CQED system with multiple two-level emitters (TLE) coupled to a single optical mode. (b)  Collective effects give rise to enhanced light-matter interactions, which result, for example, in a frequency splitting $G=g\sqrt{N}$ between the two lowest polariton modes formed by a superposition of a single photon and the collective state $|S\rangle$. (c) Two ground states $|0\rangle$ and $|1\rangle$ of an atom can be coupled to a cavity via Raman transitions involving additional laser fields with Rabi frequency $\Omega$. This technique can be used to engineer effective Dicke models with tunable parameters $G$, $\omega_c$ and $\omega_a$~\cite{dimer07}.  (d) Above a certain coupling strength, the ground state of the Dicke model undergoes a phase transition from a normal phase without any photons to a superradiant phase with a non-vanishing field expectation value, $\langle \hat a\rangle\neq 0$. At the same time the emitters are polarized along the $x$-direction, as indicated by the Bloch-sphere representation of the collective spin state.\label{fig:Collective}}
\end{figure}

The coupling of multiple two-level emitters to a single radiation mode, as studied in the context of CQED (see Fig.~\ref{fig:Collective}a), results in many nontrivial phenomena associated with the collective nature of  light-matter interactions.  In its simplest form, such a scenario is described by the Dicke model, which for a set of $N$ emitters with ground state $|g\rangle$ and excited state $|e\rangle$ reads
\begin{equation}\label{eq:HDickeModel}
\hat{H}_{\rm DM}= \hbar \omega_c \hat{a}^\dag \hat{a} + \hbar \omega_a \sum_{i=1}^N \hat{\sigma}_+^i \hat{\sigma}_-^i + \hbar g \sum_{i=1}^N (\hat{a}+\hat{a}^\dag)(\hat{\sigma}_+^i +\hat{\sigma}_-^i).
\end{equation}
Here, $\hat{a}$ is the bosonic annihilation operator for the optical mode and $\hat{\sigma}_+=(\hat{\sigma}_-)^\dag =|e\rangle\langle g|$. The  first two terms  in Eq.~(\ref{eq:HDickeModel}) describe the bare energies of the optical mode and the two-level systems, respectively, where $\omega_c$ is the cavity and $\omega_a$ the atomic frequency.  The last term accounts for the dipole-field interaction, where $g$ denotes the coupling strength between a single emitter and a single photon.

Under regular conditions, the light-matter coupling is  typically small compared to the absolute photonic energy scales such that energy non-conserving terms $\sim \hat{a} \hat{\sigma}_-^i, \hat{a}^\dag \hat{\sigma}_+^i$ can be neglected and the total number of excitations  in the system is conserved. However, due to symmetry, a single photon can only excite the fully symmetric state
\begin{equation}
|S\rangle = \frac{1}{\sqrt{N}}  \sum_{i=1}^N  |g_1\rangle|g_2\rangle \dots |e_i\rangle \dots |g_N\rangle,
\end{equation}
where the matter excitation is evenly distributed among all the emitters (see Fig.~\ref{fig:Collective}b). Importantly, the effective coupling $G=\sqrt{N} g$ between a photon and this symmetric state $|S\rangle$ is collectively enhanced by the number of emitters, while all other excited states with different symmetry are completely decoupled from the cavity.  The existence of such super- and subradiant states gives rise to various collective phenomena, which on the one hand lead to enhanced light-matter interactions or a rapid superradiant decay of ensembles of emitters~\cite{dicke54}. On the other hand, the decoupled states can be exploited for suppressing the decay of excitations, for example, for quantum memory applications~\cite{putz14,asenjo17}.

By increasing the number of dipoles or by reducing the mode volume, one eventually reaches the so-called \emph{ultrastrong} coupling regime, where the collective coupling strength $G$ becomes comparable to the photon frequency $\omega_c$. In this regime, the number of excitations is no longer conserved and even in the ground state a strong hybridization between virtual photons and matter excitations can occur. Such coupling conditions have recently become accessible in solid-state CQED systems, where dense ensembles of dipoles can be coupled to confined electromagnetic modes in the optical or THz regime~\cite{todorov10}. In parallel, it has been shown that the Dicke model can also be implemented as an effective model with cold atomic gases~\cite{ritsch13}, where the resulting ratio $G/\omega_c$ can be tuned, more or less at will, via Raman control schemes~\cite{dimer07} (see Fig.~\ref{fig:Collective}c). Indeed, by using such techniques, the long-standing prediction of a superradiant phase transition in the Dicke model has been experimentally confirmed for the first time~\cite{baumann10} (see Fig.~\ref{fig:Collective}d). Since in such a setup photons constantly leak out of the cavity, this transition is a prototype example of a non-equilibrium phase transition, which occurs in the steady state of a driven and dissipative quantum system. Compared to conventional phase transitions studied in solid-state physics, the phenomenology and universal features of their non-equilibrium counterparts is still not well understood. In this context, the combination of atomic control schemes with strongly coupled nanophotonic systems provides an interesting setting to investigate steady states, phase transitions, as well as relaxation phenomena~\cite{schutz14} in open quantum many-body systems.


\subsection{Photonic quantum simulators}

Early theoretical works in photonic quantum simulations have discussed the possibilities of attaining strongly correlated states of light in coupled resonator arrays. In this case, each resonator was interacting with a single two level system,  which could be either a real or artificial atom for the case of superconducting circuits \cite{noh17}. The early proposals suggested the simulation of the Mott to the superfluid phase transition of light in a one dimensional lattice \cite{angelakis07}, Bose-Hubbard dynamics \cite{hartmann06}, followed soon after by mean field approaches in two dimensions \cite{greentree06}.  Many  other works followed, extending these initial results to a large family of many-body phenomena including simulating artificial gauge fields in the presence of interactions for the fractional quantum Hall physics, effective spin models, and topological transport of quantum states~\cite{angelakis17}.  

\begin{figure}[h!]
\centering
\includegraphics[width=0.9\columnwidth]{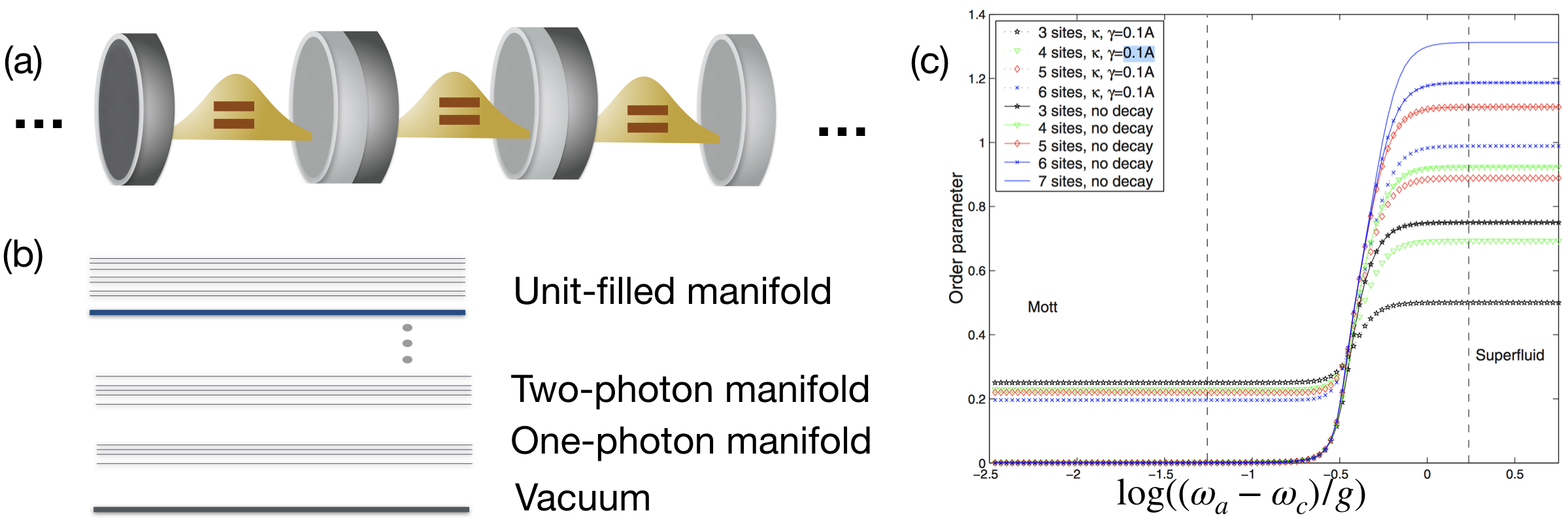}
\caption{The Jaynes-Cummings-Hubbard model. (a) A sketch of a coupled cavity array, implementing the Jaynes-Cummings-Hubbard model. (b) Energy spectrum of the JCH model. (c) The order parameter $Var(N_i)$ of the lowest-energy state in the unit-filled manifold as a function of detuning $(\omega_a-\omega_c)/g$ for 3-7 sites with and without decay. The order parameter exhibits a jump from zero to a finite value, corresponding to the Mott and the superfluid phase, respectively. The transition gets a sharper width as the system's size is increased as expected from quantum phase transition (reprinted Fig. 2 with permission from Ref.~\cite{angelakis07}. Copyright (2007) by the American Physical Society).\label{fig:mott}}
\end{figure}

These photonic resonator arrays  are  described by what is now known as the Jaynes-Cummings-Hubbard Hamiltonian (see Fig.~\ref{fig:mott}a), in analogy with the Bose-Hubbard Hamiltonian for the case of bosons on lattices, 
\begin{equation}\label{eq:HJCH}
\hat{H}_{\rm JCH} =   \hbar \sum_{j=0}^{L-1}\left[ \omega_a\hat{\sigma}^+_j\hat{\sigma}^-_j + \omega_c \hat{a}^\dagger_j \hat{a}_j +  g (\hat{a}^\dagger_j \hat{\sigma}^-_j + \hat{a}_j\hat{\sigma}^+_j)\right] - \hbar J\sum_{j=0}^{L-2}\left(\hat{a}^\dagger_{j} \hat{a}_{j+1}+{\rm H.c.}\right).
\nonumber
\end{equation}
Here, $L$ is the number of lattice sites and $J$ is the hopping strength of photons between two adjacent cavities.  At each site $j$, the first three terms in Eq.~(\ref{eq:HJCH}) describe the Jaynes-Cummings like interaction between a single two-level system with lowering (raising) operators $\hat \sigma_-^j$ ($\hat \sigma_+^j$) and a single photonic mode with bosonic annihilation (creation) operator $\hat{a}_j$ ($\hat{a}^\dagger_j$).
 $\hat{H}_{\rm JCH}$ commutes with the total number of excitations $\hat{N}=\sum_j\hat{N}_j$, where 
\begin{equation}
\hat{N}_j=\hat{a}^\dagger_j\hat{a}_j+\hat{\sigma}^+_j\hat{\sigma}^-_j. 
\nonumber
\end{equation}
Hence, the eigenenergies of $\hat{H}_{\rm JCH}$ are grouped into manifolds labeled by the filling factor $\bar{n}=\langle \hat{N} \rangle/L$, where $\langle ...\rangle$ denotes an expectation value over a given eigenstate. 

To study the many-body phases of $\hat{H}_{\rm JCH}$, let us consider the lowest energy state $|G\rangle_{\bar{n}=1}$ in the unit-filled manifold. At resonance and $g\gg J$, photon blockade prevents two photonic excitations entering the same cavity, effectively switching off the hopping process and leading to the Mott-like ground state, i.e., 
\begin{equation}
|G\rangle_{\bar{n}=1}=|1,-\rangle\otimes|1,-\rangle...\otimes|1,-\rangle.
\nonumber
\end{equation}
Here $|1,-\rangle=(|e,0\rangle-|g,1\rangle)/\sqrt2$ is the eigenstate of the local Jaynes-Cummings Hamiltonian when the cavity is resonant with the local atom.
This state can be prepared by sending a photon with $\pi/2$ pulse at frequency $\omega_a-g$ to each cavity-atom system. To shift to the superfluid phase, one needs to adiabatically increase either the hopping $J$ or the detuning $\omega_a-\omega_c$. In both cases, photon blockade will be suppressed and the system in the extreme limit of zero hopping will effectively be described by the linear tight-binding model 
\begin{equation}
\hat{H}_{\rm JCH} \approx - J\sum_j\left(\hat{a}^\dagger_{j} \hat{a}_{j+1}+{\rm H.c.}\right),
\nonumber
\end{equation}
with $|G\rangle=(\hat{a}^\dagger_{k=0})^N|000...\rangle$ as the ground state assuming $N$ excitations, which is a product state in the momentum space, not in position space as in the Mott phase.
The transition can be probed by measuring the fluctuation of the number of excitations, \textit{i.e.}, 
\begin{equation}
Var({N}_i)=\sqrt{\langle\hat{N}_j^2\rangle-\langle \hat{N}_j\rangle^2}. 
\nonumber
\end{equation}
 
Recent experimental developments of photonic lattices in different platforms ranging from semiconductor cavities and plasmonic lattices~\cite{vakevainen14} to  circuit QED  for microwave photons allowed the experimental implementation of a variety of effects. The most prominent include the simulation of a dissipative phase transition in a chain of 72 coupled nonlinear resonators \cite{fitzpatrick17} and chiral states of light using time-dependent approaches \cite{roushan16}. More recently, a quantum simulation of the many-body localization transition using interacting microwave photons in a quasi-periodic potential was implemented in a one dimensional circuit QED  lattice \cite{roushan17}. This was made possible by directly measuring statistics of eigenenergies and spreading of energy eigenstate exploiting a novel spectroscopy technique.

In conclusion, photonic quantum simulators, especially those developed in superconducting circuits provide a new promising avenue for quantum simulation due to availability of local control and measurement tools.  Experimental systems with 50-100 lattice sites are under way as well as efforts towards  two-dimensional setups. The latter will enable the controlled study of non-trivial condensed matter phenomena, such as the fractional quantum Hall effect, which involves both gauge fields and interactions and still evades other simulation approaches. Possibilities to use these analog photonic simulators in other areas are also under way, in particular also in connection with recent developments in implementing approximate algorithms for quantum chemistry and quantum machine learning. These developments motivate new experimental techniques and theoretical frameworks to maintain such controllability while increasing the systems size and will shed on many non-trivial effects in equilibrium and  driven dissipative  quantum many-body systems. 

\subsection{Nano-optomechanics}

Electromagnetic radiation and motion can interact with one another by means of the radiation pressure interaction~\cite{aspelmeyer14}. This coupling is exploited in the field of optomechanics, which explores the interaction of optical or microwave radiation with micro- or nanomechanical resonators. Nanophotonic structures can concentrate the electromagnetic energy density, confine both light and sound within a small region of space~\cite{chan12}, and thus substantially enhance this interaction whilst minimising mechanical and optical losses. The paradigmatic optomechanical interaction takes the form of a position-dependent frequency shift of the electromagnetic radiation, modelled through an interaction Hamiltonian $\hat{H}_\mathrm{INT}$ of the form
\begin{equation}
\hat{H}_\mathrm{INT}=\hbar g_{\rm om}\,\hat{a}^\dagger\hat{a}\,\hat{x}.
\end{equation}

\begin{figure}[h]
\center\includegraphics[width=0.8\columnwidth]{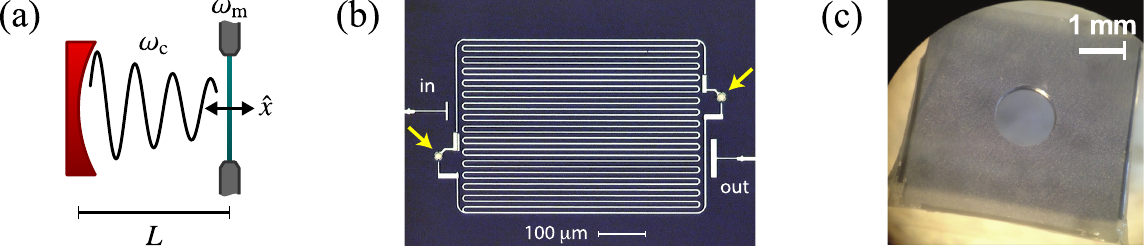}
\caption{(a)~The prototypical optomechanical system, consisting of a single mode of the electromagnetic field in a cavity of length $L$ and resonance frequency $\omega_\mathrm{c}$, one of whose end mirrors is allowed to move. The motion of this mirror is characterised by its oscillation frequency $\omega_\mathrm{m}$ and position operator $\hat{x}$. Further details are in the text. (b)~A realisation of an optomechanical system with two moving elements as a superconducting $LC$ circuit (reprinted Fig.~1a with permission from Ref.~\cite{ockeloen-korppi18}. Copyright (2018) by the Springer Nature Publishing Group). (c)~A macroscopic membrane, which may be placed \emph{inside} an optical cavity to realise the optomechanical Hamiltonian (reprinted Fig. 8 with permission from Ref.~\cite{kralj17}. Copyright (2017) by the Institute of Physics Publishing Ltd.).\label{fig:OM}}
\end{figure}

Here $g_{\rm om}$ quantifies the interaction strength between a single photon and the motion, $\hat{a}$ annihilates a photon in the electromagnetic field mode in question, and $\hat{x}$ is the dimensionless position operator for the mechanical resonator in units of its zero-point fluctuations. The coupling strength $g_{\rm om}$ is a sensitive function of the geometry of the system, and increases with a decrease in the electromagnetic mode volume or the mass or frequency of the mechanical resonator. Indeed, for a one-dimensional electromagnetic cavity of length $L$, the interaction strength between a field mode with frequency $\omega_\mathrm{c}$ and a mechanical resonator of mass $m$ and frequency $\omega_\mathrm{m}$ reads
\begin{equation}
g_{\rm om}=\sqrt{\frac{\hbar}{m\omega_\mathrm{m}}}\frac{\omega_\mathrm{c}}{L}.
\end{equation}
When $g_{\rm om}$ is very small on the scale of the other frequencies of the system, it is possible to amplify its effects by operating at large photon numbers. This comes at the cost of linearising the dynamics, yielding an interaction of the form
\begin{equation}
\hat{H}_\mathrm{INT}=G\bigl(\hat{a}+\hat{a}^\dagger\bigr)\hat{x},
\end{equation}
where $G \gg g_{\rm om}$ is the linearised optomechanical interaction strength. Reaching the single-photon strong coupling regime, where $g_{\rm om}$ becomes comparable to the decay rate of the electromagnetic cavity and the oscillation frequency of the mechanical resonator, is an outstanding goal of the field. New behaviour, such as novel blockade effects~\cite{rabl11}, are predicted to occur in this regime. It has been suggested~\cite{xuereb12} that the use of multiple mechanical elements interacting with the same electromagnetic field mode may enhance the optomechanical interaction with the collective mechanical motion and give access to this regime.

This parametric coupling between light and motion enables many possibilities, including ground-state cooling of vibrational modes, entanglement between light and motion or between two mechanical resonators, mechanical storage of quantum information, coherent conversion between optical and microwave signals~\cite{tian14}, switches actuated by individual photons, and non-reciprocal photonic devices~\cite{bernier17, barzanjeh17, ruesink18}. It also makes possible sensing and readout of minute displacement forces and fields~\cite{schilling16} surpassing the standard quantum limit. The linearised optomechanical coupling introduced earlier also makes possible the time-dependent control of micro- and nanomechanical devices, through the linear dependence of $G$ on the amplitude of the electromagnetic field. Application of nano-optomechanics to real-world scenarios in the quantum regime, therefore, requires the development of efficient quantum control techniques and quantum information-processing protocols. This problem is compounded in the case of systems with multiple mechanical resonators, where inevitable manufacturing differences between resonators necessitate novel protocols. In the single-photon strong coupling regime and when nonlinear dynamics are significant, new theoretical tools are required to model the effects of radiation pressure at the few-photon level. Recent experiments have demonstrated the preparation of entangled mechanical states in the linearised regime~\cite{ockeloen-korppi18, riedinger18}, but the conditional preparation of more generic non-classical mechanical states is outstanding. The theory of systems with multiple mechanical resonators, and optomechanical lattices in the limit of very large numbers of resonators, is still relatively unexplored. Other outstanding theoretical challenges include defect--phonon interactions in nanobeams for enhanced phonon nonlinearities; and the modelling of the mechanical properties of truly nanoscopic systems such as carbon nanotubes, and graphene and other two-dimensional materials.

Nano- and optomechanical phenomena are ubiquitous and can be observed in a vast array of photonic, nanophotonic, optical, and microwave systems. Achieving strong interactions requires optimisation of both photonic and mechanical properties. Significant work is currently taking place in the engineering of nanoscale optomechanical systems with strong optomechanical coupling and low optical absorption, phononic shields and optimise materials that allow to obtain ultrahigh mechanical quality factors, uniform arrays of mechanical resonators to create optomechanical lattices, and integrated nanophotonic and microwave optomechanical systems for coherent optics-to-microwave conversion. A closely related branch of work deals with coupling mechanical motion to individual quantum emitters~\cite{lee17}. Of current interest here is the near-field coupling of carbon nanotubes and graphene mechanical resonators to optical resonators and single quantum emitters, as well as diamond optomechanical systems coupled to nitrogen- and silicon-vacancy defect centers.

Mechanical resonators of interest in the field of nano-optomechanics tend to have oscillation frequencies in the region of MHz to a few GHz. Under most circumstances, this makes such systems highly susceptible to thermal noise and decoherence processes. The quantum control of individual mechanical resonators has been demonstrated in cryogenic environments. Any move to real-world optomechanical technologies, however, requires devices that must operate at, or at least close to, room temperature. One of the significant open challenge hindering their uptake is the large-scale manufacturing of optomechanical devices with low disorder, which is a necessary stepping stone to the scalability of nano-optomechanical systems beyond one or very few elements, as well as the complementary metal-oxide-semiconductor (CMOS) compatible manufacture of optomechanical devices. Solving these challenges would open the door to the widespread use of nano-optomechanical technologies for basic science, information and communication technology, as well as sensing and metrology.

\section{Technological and industrial outlook}

In recent years, several companies have emerged or have expanded their activities in the field of NQO, see for example the COST
Action MP1403 industry partners at http://www.cost-nqo.eu/industry. Many companies have strong ties with academic research laboratories, highlighting the importance of academia/industry collaboration. This also emphasizes the potential of NQO for creating viable and sustainable companies by training highly skilled personnel and by developing risky technologies in an academic environment. Here we discuss several aspects of NQO that have already made an impact in industry. A detailed analysis focused on quantum communication and sensing is presented in a recent market research study~\cite{MRS19}.

\subsection{Single-photon sources}

The deterministic generation of pure single photons is a now highly sought-after technology for many research laboratories. The probabilistic generation of single photons from spontaneous parametric processes is currently severely hindering the progress of optical quantum information processing and quantum communication, and NQO offers several approaches based on compact and integrated sources such as QDs, color centers, quantum light-emitting diodes (LEDs) and organic molecules that have the potential to yield high-rate, high-quality and, importantly, deterministic single-photon emission. Some emerging companies active in NQO are now offering single-photon sources that can be integrated in research laboratories. This is a first step that will lead to the generation of turnkey and user-friendly systems that will further facilitate dissemination by the academic and industrial communities. In terms of performance, the immediate needs are: Self-contained turnkey system (in a cryostat if necessary), large coupling efficiency into an optical fiber to reach deterministic generation inside an optical fiber, high-rate and deterministic emission to demonstrate a gain over spontaneous generation, electrical pumping, telecom wavelengths, deterministic and high-quality sources of entangled photon pairs, reproducible and scalable nanofabrication technologies (in-situ optical and electron-beam lithography).  The industry is now in a position to address these different aspects, which will lead to the development of performance standardization procedures.

\subsection{Single-photon detectors}

Superconducting nanowire single-photon detection, which offers unrivalled efficiency and detection rate, low noise and broadband operation, is currently a very active area of NQO in industry, with several companies offering systems and solutions. Turnkey operation has been reached, and users can now acquire fully operation systems that can be easily integrated in academic and industrial environments. While academic laboratories remain a strong acquirer of this technology, one challenge is now to disseminate the technology further in industrial applications and gain large acceptance. Some aspects that require further development are: Larger collection areas for free-space coupling or multimode fiber coupling, higher detection speed (towards GHz
counting rate) in a cost-effective way, mid-infrared sensitivity (up to 5 $\mu$m), integration of high-performance detectors in quantum photonics platforms. Standardization of the performances will also play an important role in widening the acceptance of this technology in industry.

\subsection{Photonic integration}

Integration of NQO technologies in optical circuits is an important and timely objective for several industrial companies active in integrated optics. Indeed, companies and research laboratories now aim to leverage from the developments in the ``optics-on-a-chip'' field to start integrating single-photon sources, detectors and processing capabilities on the quantum photonics platform. Several materials are being pursued, such as GaAs, SOI, lithium niobate, silica and diamond. A unified approach involving both academics and industry is essential to converge rapidly. Targeting specific applications such as quantum key distribution on a chip using deterministic and high-rate single-photon
sources could achieve this. Another important application is computation tasks, such as Boson sampling on a chip, in which the combination of multiphoton generation, processing and detection could lead to the demonstration of quantum supremacy. The role of industry in this line of work is to find ways to develop rapid, performing and cost-effective ways to create integrated circuits. Fortunately, this progress will go together with that in integrated optics, which will be highly beneficial.

\subsection{Single-spin sensing}

Single-spin sensing has rapidly evolved in the last few years, and several companies have emerged in the area. These companies leverage from nanofabrication techniques and established diamond providers to explore the possibility of sensing with nanometre-scale precision and unprecedented sensitivity. Applications range from failure analysis of electronic systems (through magnetic or electric field sensing) to life sciences, where an understanding of the intracell mechanisms is key to understanding subcellular processes and to early diagnosis. One important challenge is therefore to disseminate the technology in areas outside of the quantum physics/nanotechnology sector. This inevitably requires a consolidation of the technology in a rugged and turnkey system that can be adopted by many end users with diverse backgrounds. This will be combined with academic innovation, and the synergy between companies and laboratories will allow rapid development cycles.

\subsection{Numerical modelling}

Numerical modeling has played a crucial role in the development of nano-optics in research laboratories. Several companies are active in this field and are pursuing the development of software that will be useful for NQO as well. This could require adapting the numerical methods to take quantum effects into account, which would lead to a unique tool for developing NQO processes. Possible application areas cover fundamental quantum many-body physics, multidimensional coherent spectroscopy modelling tools, etc.

\subsection{Enabling technologies}

The core technology in NQO relies on more generic technologies such as lasers, high- speed electronics and cryogenics. The development of user-friendly tools exploiting NQO technologies will greatly benefit from partnerships with the generic technology companies. Such partnerships will be essential to disseminate NQO technologies, and will allow both sides to expand their market.


 
\section*{Acknowledgments}

This article is based upon work from COST Action MP1403 ``Nanoscale Quantum Optics,'' supported by COST (European Cooperation in Science and Technology). The authors would like to acknowledge input from D. Chang, V. Giesz, S. K\"uck, R. Oulton, and C. Sibilia. T. Durt and B. Kolaric acknowledge the help of Mrs. B. Bokic in editing images for the section Nanoscale Quantum Coherence. A. Xuereb acknowledges funding by the European Union’s Horizon 2020 research and innovation programme under grant agreement No 732894 (FET Proactive HOT) and wishes to thank G. Di Giuseppe, N. Kralj, E. Serra and D. Vitali for providing Fig.~\ref{fig:OM}c

\bibliography{manuscript}

%
%
%

\end{document}